\documentclass[twocolumn]{autart}

\usepackage{cite,comment}
\usepackage{mathrsfs}
\usepackage{graphicx}
\usepackage{graphics}
\usepackage{amssymb}
\usepackage{amsmath,mathtools}
\usepackage{color}
\usepackage{xspace}
\usepackage{algpseudocode}
\usepackage{bbm}
\usepackage{comment}
\usepackage{algorithmicx}
\usepackage[caption=false]{subfig}
\usepackage{psfrag}
\usepackage{url}
\usepackage{float}
\usepackage{caption}
\usepackage{soul}

\usepackage{rotating}
\usepackage{chngcntr}
\usepackage{apptools}
\AtAppendix{\counterwithin{thm}{section}}
\usepackage{graphicx}
\usepackage{graphics}
\usepackage{amssymb}
\usepackage{amsmath}
\usepackage{mathtools}
\usepackage{color}
\usepackage{xspace}
\usepackage{algpseudocode}
\usepackage{bbm}
\usepackage{comment}
\usepackage{algorithmicx}
\usepackage{subfig}
\usepackage{psfrag}
\usepackage{tikz}
\usetikzlibrary{shapes,arrows}
\usetikzlibrary{positioning}

\usepackage{rotating}
\usepackage{chngcntr}
\usepackage{apptools}
\usepackage{listings}
\AtAppendix{\counterwithin{thm}{section}}
\usepackage{graphicx}
\usepackage{graphics}
\usepackage{amssymb}
\usepackage{amsmath}
\usepackage{color}
\usepackage{xspace}
\usepackage{algpseudocode}
\usepackage{bbm}
\usepackage{comment}
\usepackage{algorithmicx}
\usepackage{subfig}
\usepackage{psfrag}
\usepackage{tikz}
\usetikzlibrary{shapes,arrows}
\usetikzlibrary{positioning}

\usepackage{enumitem}
\usepackage{graphicx}               
\usepackage{amsmath,amssymb}

\usepackage{graphics}
\usepackage{amssymb}
\usepackage{amsmath}
\usepackage{color}
\usepackage{xspace}
\usepackage{algpseudocode}
\usepackage{bbm}
\usepackage{comment}
\usepackage{algorithmicx}
\usepackage{subfig}
\usepackage{psfrag}
\usepackage{soul}
\usepackage{tikz}
\usetikzlibrary{petri}
\usetikzlibrary{shapes,arrows}
\usetikzlibrary{positioning}

\usepackage{cancel}

\tikzstyle{block}=[draw opacity=0.7,line width=1.4cm]

\newcommand{\oprocendsymbol}{\hbox{$\bullet$}}
\newcommand{\oprocend}{\relax\ifmmode\else\unskip\hfill\fi\oprocendsymbol}

\newcommand{\longthmtitle}[1]{\mbox{}\textit{{(#1):}}}

\newcommand{\VV}{\mathcal{V}}
\newcommand{\EE}{\mathcal{E}}
\newcommand{\GG}{\mathcal{G}}

\newcommand{\iI}{\vect{\mathsf{I}}}

\newcommand{\lL}{\vect{\mathsf{L}}}
\newcommand{\rR}{\vect{\mathsf{R}}}
\newcommand{\rr}{\vect{\mathsf{r}}}

\newcommand{\pPi}{\vect{\mathsf{\Pi}}}

\newcommand{\xs}{\vect{\mathsf{x}}^{\star}}

\newcommand{\xsi}{{\vectsf{x}}^{i\star}}

\newcommand{\bs}{\mathsf{b}}

\newcommand{\nus}{{\vectsf{\nu}}^{\star}}

\newcommand{\real}{{\mathbb{R}}}
\newcommand{\reals}{{\mathbb{R}}}
\newcommand{\realpositive}{{\mathbb{R}}_{>0}}
\newcommand{\realnonnegative}{{\mathbb{R}}_{\ge 0}}
\newcommand{\integerpositive}{{\mathbb{Z}}_{>0}}

\newcommand{\eps}{\epsilon}

\newcommand{\argmin}{\operatorname{argmin}}
\newcommand{\rank}{\operatorname{rank}}

\newcommand{\until}[1]{\in\{1,\dots,#1\}}
\newcommand{\vect}[1]{\boldsymbol{\mathbf{#1}}}
\newcommand{\vectsf}[1]{\boldsymbol{\mathbf{\mathsf{#1}}}}
\newcommand{\Bvect}[1]{\bar{\boldsymbol{\mathbf{#1}}}}

\newcommand{\dvect}[1]{\dot{\vect{#1}}}

\newcommand{\Diag}[1]{\operatorname{Diag}(#1)}

\newcommand{\SUM}[2]{\sum_{#1}^{#2}}
 \newcommand{\boxend}{\hfill \ensuremath{\Box}}

\newcommand{\bunderline}[1]{\underline{#1\mkern-2mu}\mkern2mu }

\newtheorem{assump}{Assumption}[section]

\newtheorem{thm}{Theorem}[section]
\newtheorem{prop}{Proposition}[section]
\newtheorem{rem}{Remark}[section]

\newtheorem{lem}{Lemma}[section]


\tikzstyle{block}=[draw opacity=0.7,line width=1.4cm]
\DeclareMathAlphabet{\mathpzc}{OT1}{pzc}{m}{it}
\definecolor{CranJ}{cmyk}{0,0.69,0.54,0.04} 
\definecolor{PinkJ}{cmyk}{0,0.71,0.43,0.12} 
\definecolor{Cran}{cmyk}{0,0.73,0.41,0.29} 
\definecolor{VRed}{cmyk}{0,0.75,0.25,0.2} 
\definecolor{ORed}{cmyk}{0,0.75,0.75,0} 
\definecolor{CBlue}{cmyk}{1,0.25,0,0} 




\allowdisplaybreaks
\begin{document}

\begin{frontmatter}
   \runtitle{Distributed optimal resource allocation}
  
   \title{Cluster-based Distributed Augmented Lagrangian Algorithm for a Class of Constrained Convex Optimization Problems \thanksref{footnoteinfo}} 
  
  \thanks[footnoteinfo]{Corresponding author: H. Moradian}
      \author[Paestum]{Hossein Moradian}\ead{hmoradia@uci.edu}\quad
  \author[Paestum]{Solmaz S. Kia}\ead{solmaz@uci.edu} 
\thanks[footnoteinfo]{This work was supported by NSF, United States of America, CAREER  award ECCS-1653838. A preliminary  version of this paper is presented in~\cite{SSK:17acc}.}

  \address[Paestum]{Department of Mechanical and Aerospace
    Engineering, University of California, Irvine}
  \begin{abstract}
     We propose a distributed solution for a constrained convex optimization problem over a network of clustered agents each consisted of a set of subagents. The communication range of the clustered agents is such that they can form a connected undirected graph topology. The total cost in this optimization problem is the sum of the local convex costs of the subagents of each cluster. We seek a minimizer of this cost subject to a set of affine equality constraints, and a set of affine inequality constraints specifying the bounds on the decision variables if such bounds exist. We design our distributed algorithm in a cluster-based framework which results in a significant reduction in communication and computation costs. Our proposed distributed solution is a novel continuous-time algorithm that is linked to the augmented Lagrangian approach. It converges asymptotically when the local cost functions are convex and exponentially when they are strongly convex and have Lipschitz gradients.
    Moreover, we use an $\eps$-exact penalty function to address the inequality constraints and derive an explicit lower bound on the penalty function weight to guarantee convergence to $\eps$-neighborhood of the global minimum value of the cost. A numerical example demonstrates our results.
  \end{abstract}
\end{frontmatter}

\section{Introduction}
\begin{figure}[t]
  \centering
\includegraphics[trim=10pt 15pt 10pt 10pt, clip,scale=.34]{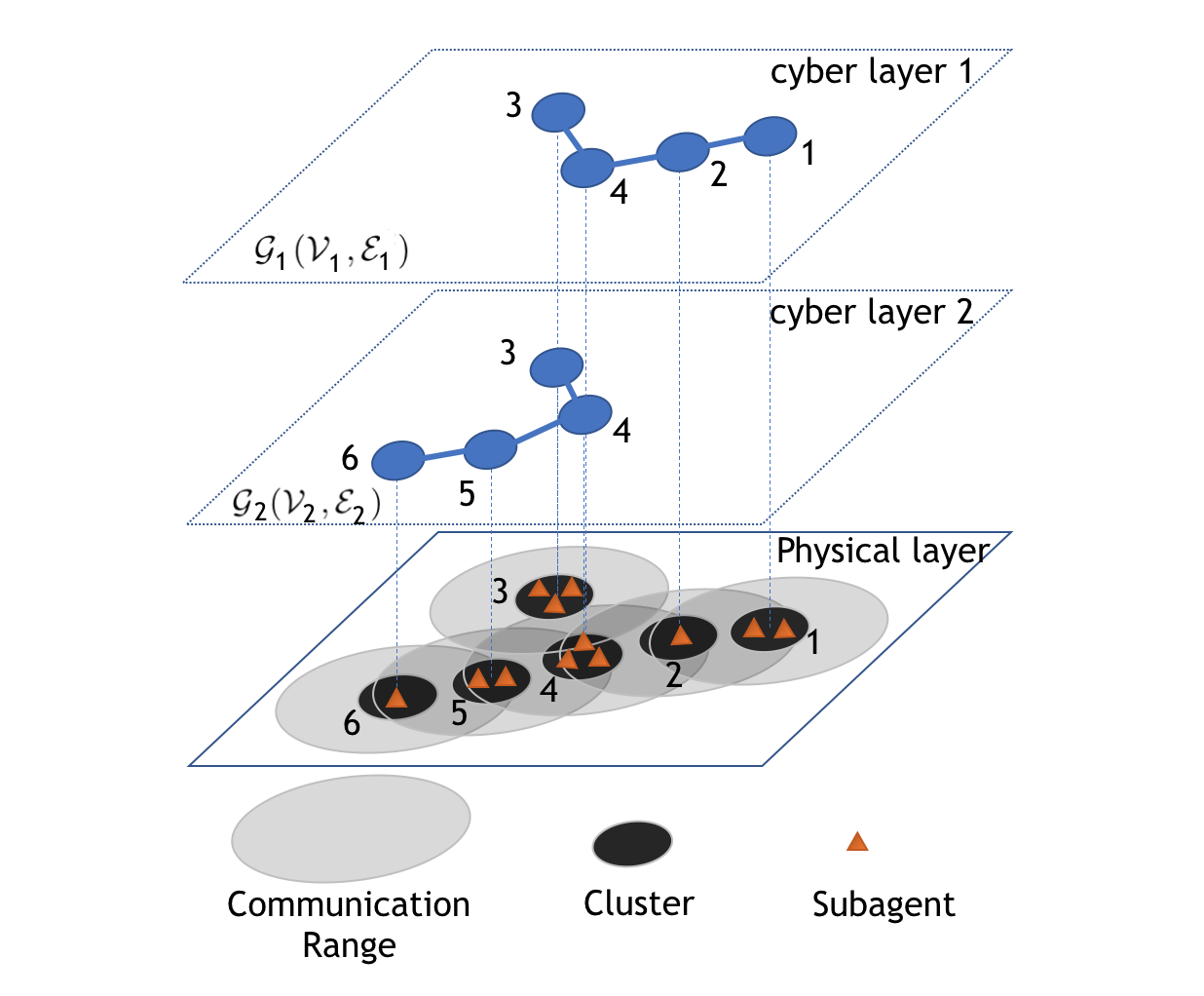} 
      \vspace{-2pt}
  \caption{ {\scriptsize A group of clustered agents (generators) with undirected connected graph topology aim to solve $\xs=\arg\min_{\vect{x}\in
    \reals^{12}}\sum\nolimits_{i=1}^6 f^i(\vect{x}^i)$, subject to $[1~ 1]\vect{x}^1+\,\vect{x}^2+[0.5~ 0.5~ 0.5]\vect{x}^3+\,[1\,\,1\,\,1]\vect{x}^4=450$, 
 $[0.5~ 0.5~ 0.5]\vect{x}^3+[1~ 1]\,\vect{x}^5+\vect{x}^6=700$, and 
$\bunderline{\mathsf{x}}^i_l\!\leq\! x^i_l\!\leq\!\bar{\mathsf{x}}^i_l,\quad\quad i\!\in\!\mathbb{Z}_1^6,\quad l\!\in \!\mathbb{Z}_1^{n^i}$ in a distributed manner. Here, $f^i(\vect{x}^i)=\SUM{l=1}{n^i}f^i_l(x^i_l)$, where  $f^i_l(x^i_l)=\alpha^i_lx^i_l\,\!^2+\beta^i_lx^i_l+\gamma^i_l$. In the physical layer plot, a cluster agent can communicate with another cluster if it is inside the other cluster's communication disk. To solve this optimal resource allocation problem in a distributed manner, we form subgraphs $\GG_1(\VV_1,\EE_1)$ and $\GG_2(\VV_2,\EE_2)$, which are associated, respectively, with the first and the second equality constraints. Here, $\VV_1=\{1,2,3,4\}$ and $\VV_2=\{3,4,5,6\}$. Agent $4$ acts as a connectivity helper node in $\GG_2$. A solution to this problem using our proposed algorithm is given in~section~\ref{sec::num}.
} }\label{fig::network}
\end{figure}

\vspace{-0.13in}
We consider a group of $N$ clustered agents $\VV=\{1,\cdots,N\}$ with communication and computation capabilities, whose communication range is such that they can form a connected undirected graph topology, see Fig.~\ref{fig::network}. These agents aim to solve, in a distributed manner, the optimization problem 
\begin{subequations} \label{eq::prob_def}
\begin{align}
  \xs&=\arg\min_{\vect{x}\in\real^m} \,\,\sum\nolimits_{i=1}^N f^i(\vect{x}^i),~~\text{subject~to~} \\
    &\!\!\!\!\!\!\![\vectsf{w}^1]_j\vect{x}^1\!\!+\cdots+[\vectsf{w}^N]_j\vect{x}^N\!\!-\bs_j\!=\!0,\,\,
    j\in\{1,\cdots,p\},\label{eq::prob_def-equal} \\
   & \bunderline{\mathsf{x}}^i_l\leq{x}^i_l,\quad  \,\,\,l\in\underline{\mathcal{B}}^i\subseteq\{1,\cdots,n^i\},\quad i\in\VV,\label{eq::prob_def-box1}\\
   & {x}^i_l\leq \bar{\mathsf{x}}^i_l,\quad \,\,\,l\in\bar{\mathcal{B}}^i\subseteq\{1,\cdots,n^i\},\quad  i\in\VV,\label{eq::prob_def-box2}
\end{align}
\end{subequations}
where 
$f^i(\vect{x}^i)=\sum\nolimits_{l=1}^{n^i} f_l^i(x_l^i).$ In this setting, each agent $i\in\VV$ is a \emph{cluster} of local `subagents'  $l\until{n^i}$ whose decision variable is $\vect{x}^i=[x^i_1,\cdots,{x}^i_{n^i}]^\top\in\real^{n^i}$. The weighting factor matrix $\vectsf{w}^i\in\real^{p\times n^i}$ of each agent $i\in\VV$ is only known to the agent $i$ itself. Moreover, $\bunderline{\mathsf{x}}^i_l,\bar{\mathsf{x}}^i_l\in\real$, with $\bunderline{\mathsf{x}}^i_l<\bar{\mathsf{x}}^i_l$, are respectively the 
lower and upper bounds on the $l^\text{th}$ decision variable of agent $i\in\VV$, if such a bound exists. In a distributed solution, each agent $i\in\VV$ should obtain its respective component of $\vectsf{x}^{\star}=[{\vectsf{x}^{1\star\top}},\cdots,{\vectsf{x}^{N\star\top}}]^\top$ by interacting only with the agents that are in its communication range. 
Problem~\eqref{eq::prob_def} explicitly or implicitly, captures various in-network optimization problems.  
One example is the optimal in-network resource allocation, which appears in many optimal decision making tasks such as economic dispatch over power networks~\cite{AJW-FW-GBS:13,AC-JC:16},  
 optimal routing~\cite{LX-M-SPB:04,RM-SL:06} 
and network resource allocation for wireless systems~\cite{JC-VKNL:12,AF-FP:14}. 
In such problems, a group of agents with limited resources, e.g., a group of generators in a power network, add up their local resources to meet a demand in a way that the overall cost is optimum for the entire network. Another family of problems that can be modeled as~\eqref{eq::prob_def} is the in-network model predictive control over a finite horizon for a group of agents with linear dynamics~\cite{SAA-KY-AHS:18,RR-GC-DG:17}. 

\vspace{-0.05in} 

 In recent years, there has been a surge in the design of distributed algorithms for large-scale in-network optimization problems. The major developments have been in the  unconstrained convex optimization setting where the global cost is the sum of local costs of the agents (see e.g.~\cite{SB-NP-EC-BP-JE:10,JD-AA-MW:12}
  for algorithms in discrete-time, and~\cite{JW-NE:11,SSK-JC-SM:15-auto, DV-FZ-AC-GP-LS:15}
 for algorithms in continuous-time). In-network constrained convex optimization problems have also been studied in the literature. For example, in the context of the power generator economic dispatch problem,~\cite{ZZ-MC:12,SK-GH:12,ADDG-STC-CNH:12} offer distributed solutions that solve a special case of~\eqref{eq::prob_def} with local quadratic costs subject to bounded decision variables and a single demand equation, $p=1$ and $\mathsf{w}^i=1$ for $i\in\VV$. Distributed algorithm design for special cases of~\eqref{eq::prob_def} with non-quadratic costs are presented in~\cite{LX-SB:06,YZ-MMZ:18,SAA-KY-AHS:18}
 in discrete-time form, and~\cite{PY-YH-FL:16,SSK:17,DD-MRJ:18,AC-JC:16-auto,AC-JC:15-tcns}
in continuous-time form. 
Except for~\cite{YZ-MMZ:18}, all these algorithms consider the case that the local decision variable of each agent $i\in\VV$ is a scalar. Moreover,
with the exception of~\cite{SSK:17,YZ-MMZ:18,SAA-KY-AHS:18}, these algorithms only
 solve~\eqref{eq::prob_def} when the equality constraint is the unweighted sum of local decision variables, i.e., $p=1$ and $\mathsf{w}^i=1$ for $i\in\VV$. Also, only~~\cite{AC-JC:16-auto} and~\cite{AC-JC:15-tcns}  consider local inequality constraints, which are in the form of local box inequality constraints on all the decision variables of the problem. Lastly, the algorithms in~\cite{LX-SB:06,AC-JC:16-auto,AC-JC:15-tcns} require the agents to communicate the gradient of their local cost functions to their neighbors. Such a requirement can be of concern for privacy-sensitive
applications.

\vspace{-0.05in}

In this paper, we propose a novel distributed algorithm to solve the optimization problem~\eqref{eq::prob_def}. We start by considering the case that $\underline{\mathcal{B}}^i=\bar{\mathcal{B}}^i=\{\}$ for $i\in\VV$, i.e., when there is no inequality constraint. For this problem, 
we propose a continuous-time distributed primal-dual algorithm. To induce robustness and also to yield convergence without strict convexity of the local cost functions, we adapt an augmented Lagrangian framework~\cite{DPB-JNT:97}.
The augmented Lagrangian method has been used in~\cite{DJ-JMFM-JX:15},~\cite{MV-JC:18}, and~\cite{YZ-MMZ:18} to improve the transient response of the distributed algorithms for, respectively, 
an unconstrained convex optimization, an online optimization, and a discrete-time constrained optimization problems. 
Different than the customary practice of using a common augmented Lagrangian penalty parameter as in~\cite{YZ-MMZ:18,MV-JC:18,DJ-JMFM-JX:15}, in our design to reduce the coordination overhead among the agents we allow each agent to choose its own penalty parameter locally. The structure of our distributed solution is inspired by the primal-dual centralized solution of~\cite{KJA-LH-HU:58} (see~\eqref{eq::saddle-aug}), where the coupling in the differential solver is in the dual state dynamics. In decentralized primal-dual algorithms, e.g.~\cite{DD-BH-NKD-MRJ:18,DD-MRJ:18,SSK:16-ifac}, the adopted practice is to give every agent a copy of the dual variables and use a consensus mechanism to make the agents arrive eventually at the same dual variable. We follow the same approach but in our design, we pay particular attention to computation and communication resource management by adopting a cluster-based approach. First, we consider the sparsity in the equality constraints and give only a copy of a dual variable to an agent if a decision variable of that agent is involved in the equality constraint corresponding to that dual variable. Then, only the cluster of the agents that have a copy of the dual variable need to form a connected graph and use a consensus mechanism to arrive at agreement on their dual variable, see Fig.~\ref{fig::network}. Next, in our design, we only assign a single copy of the dual variable to an agent $i$ regardless of how many subagents it has. We note that if we use the algorithms in~\cite{LX-SB:06,YZ-MMZ:18,SAA-KY-AHS:18,PY-YH-FL:16,SSK:17,DD-MRJ:18,AC-JC:16-auto,AC-JC:15-tcns} to solve problems where $\vect{x}^i\in\real^{n^i}$ of an agent $i\in\VV$ is a vector ($n^i>1$), we need to treat each component of the $i$ as an agent and assign a copy of a dual variable to it. Such a treatment increases the local storage, computation and communication costs of agent $i$. Our convergence analysis, 
is based on the Lyapunov and the LaSalle invariant set methods, and also the semistability analysis~\cite{WMH-VC:08} to show that our algorithm is guaranteed to converge to a point in the set of optimal decision values when the local costs are convex. When the local cost functions are strongly convex and their local gradients are globally Lipschitz the convergence guarantees of our proposed algorithm over connected graphs is exponential and can also be extended to dynamic graphs. 

\vspace{-0.05in}

To address scenarios where all or some of the decision variables are bounded in~\eqref{eq::prob_def}, we use a variation of exact penalty function method~\cite{DPB:75}, called $\eps$-exact penalty function method~\cite{MCP-SAZ:94}. Unlike the exact penalty method, this method uses a smooth differentiable penalty function to converge to the $\eps$-neighborhood of the global minimum value of the cost. The advantage of exact penalty function methods is in the possibility of using a finite penalty weight to arrive at a  practical and numerically well-posed optimization solution. However, as shown in~\cite{DPB:75,MCP-SAZ:94}, the penalty function weight is lower bounded by the bounds on the Lagrange multipliers. Since generally, the Lagrange multipliers are unknown, the bound on the penalty function weight is not known either. Many literature that use penalty function methods on distributed optimization framework generally state that a large enough value for the weight is used~\cite{WW-JW-NL-SM:17,MZ-DF-ADD:19}, with no guarantees on the feasibility of their choice. 
\cite{OM:85},~\cite{SR-MM-CJ:11},~\cite[Lemma 5.1]{AC-JC:15-tcns} and~\cite[Proposition 4]{SSK:16-ifac},, and are among few results in literature that address the problem of establishing an exact upper-bound on the size of the Lagrange multipliers, which can be used to obtain a lower bound on the size of the valid penalty function weight.  However,~\cite{OM:85} only considers problems with inequality constraints only, while~\cite[Lemma 5.1]{AC-JC:15-tcns},~\cite[Proposition 4]{SSK:16-ifac} are developed for the resource allocation problem described by~\eqref{eq::prob_def} when there exists \emph{only} one equality constraint ($p=1$) with $\mathsf{w}^i=1$, $i\in\VV$ and all the decision variables have boxed inequality. on the other hand~\cite{SR-MM-CJ:11} proposes a numerical procedure. As part of our contribution in this paper, we obtain an explicit closed-form upper-bound on the Lagrange multipliers of problem~\eqref{eq::prob_def}, which 
enables determining the size of the suitable penalty function weight for both exact and $\eps$-exact penalty function methods.

\vspace{-0.05in}
In summary, the contribution of this paper is twofold. (a) We propose a novel distributed algorithm to solve problem~\eqref{eq::prob_def}. This design uses an augmented Lagrangian approach, which, similar to the case of centralized solvers, extends the convergence guarantees of our proposed distributed algorithm to convex cost functions, as well. Our design also incorporates a cluster-based approach to reduce computational and communication costs. (b) We establish a well-defined upper-bound on the Lagrange multipliers of problem~\eqref{eq::prob_def}. This result is of fundamental importance and its impact is beyond our proposed algorithm. It is useful in identifying the value of the weight factor of exact and $\epsilon$-exact penalty functions that are used to address inequality constraints.

\vspace{-0.05in}
\section{Preliminaries}\label{sec::prelim}
\vspace{-0.13in}
Let $\reals$, $\real_{\geq 0}$, $\mathbb{Z}$, and $\integerpositive$ be, respectively, the set of real, nonnegative real, integer, and positive integer numbers. For a given $i,j\in\mathbb{Z}$, $i<j$, we define $\mathbb{Z}_{i}^j=\{x\in\mathbb{Z}\,|\,i\leq x\leq j\}$. We denote the cardinality of a set $\mathcal{A}$ by $|\mathcal{A}|$.
For a matrix $\vect{A}=[\mathsf{a}_{ij}]\in\real^{n\times m}$,  we denote its transpose matrix by~$\vect{A}^\top$,  $k^{th}$ row by $[\vect{A}]_k$, $k^{th}$ column by $[\vect{A}]^k$, and its element wise max-norm with $\|\vect{A}\|_{\max}$
. We let $\vect{1}_n$
(resp. $\vect{0}_{n}$) denote the vector of $n$ ones (resp. $n$
zeros),  $\vectsf{I}_n$ denote the $n\times n$ identity
matrix and $\pPi_n= \iI_n - \frac{1}{n}\vect{1}_n\vect{1}_n^\top$.
When clear from the context, we do not
specify the matrix dimensions. For a vector $\vect{x}\in\reals^n$ we denote the standard Euclidean and infinity norms by, respectively,  $\|\vect{x}\|\!=\!\sqrt{\vect{x}^\top\vect{x}}$ and $\|\vect{x}\|_\infty\!=\!\max{|x_i|}_{i=1}^n$. Given a set of vectors,  we use $[\{\vect{p}^i\}_{i\in\mathcal{M}}]$ to indicate the aggregate vector obtained from staking the set of the vectors $\{\vect{p}_i\}_{i\in\mathcal{M}}$  whose indices belong to the ordered set $\mathcal{M}\subset\mathbb{Z}_{>0}$. In~a network of $N$ agents, to distinguish and emphasize that a variable is local to an agent $i\in\mathbb{Z}_{1}^N$, we use superscripts, e.g., $f^i(\vect{x}^i)$ is the local function of agent $i\in\mathbb{Z}_{1}^N$ evaluated at its own local value $\vect{x}^i\in\real^{n^i}$. The $l^{th}$ element of a vector $\vect{x}^i\in\real^{n^i}$ at agent  $i\in\mathbb{Z}_{1}^N$ is denoted by $x_l^i$. Moreover, if
$\vect{p}^i\in\reals^{d^i}$ is a variable of agent $i\in\VV=\{1,\cdots,N\}$, the aggregated
$\vect{p}^i$'s of the network is the vector~$\vect{p} =[\{\vect{p}^i\}_{i\in\VV}]=
[{\vect{p}^1}^\top,\cdots,{\vect{p}^N}^\top]^\top \in \reals^{\bar{d}}$and
$\text{Blkdiag}(\vect{p})=\Big[
\begin{smallmatrix} \vect{p}^1& \vect{0}& \vect{0} \\ \vect{0}&\cdots &\vect{0}\\ \vect{0}& \vect{0}& \vect{p}^N \end{smallmatrix}\Big] \in\real^{\bar{d}\times N}$, with $\bar{d}=\sum\nolimits_{i=1}^N d^i$.  
For a differentiable function $f: \reals^d\to\reals$, $\nabla f(\vect{x})$ represents its gradient.  A differentiable function $f: \reals^d\to\reals$ is convex  (resp. $\alpha$-strongly convex, $\alpha\in\realpositive$)
over a convex set $C \subseteq \reals^d$~if and only if $
  (\vect{\mathsf{z}}-\vect{\mathsf{x}})^\top(
  \nabla f(\vect{\mathsf{z}})-\nabla f(\vect{\mathsf{x}}))\geq
  0$ (resp. $
  \alpha\|\vectsf{z}-\vectsf{x}\|^2\leq  (\vect{\mathsf{z}}-\vect{\mathsf{x}})^\top(\nabla f(\vect{\mathsf{z}})-\nabla f(\vect{\mathsf{x}}))
  $, or equivalently $\alpha\|\vectsf{z}-\vectsf{x}\|\leq  \|\nabla f(\vect{\mathsf{z}})-\nabla f(\vect{\mathsf{x}})\|$)   for all $\vect{\mathsf{x}},\vect{\mathsf{z}}\in C$. Moreover, it is strictly convex over a convex set $C \subseteq \reals^d$~if and only if $
  (\vect{\mathsf{z}}-\vect{\mathsf{x}})^\top(
  \nabla f(\vect{\mathsf{z}})-\nabla f(\vect{\mathsf{x}}))>0$.

\vspace{-0.05in}
Next, we briefly review basic concepts from algebraic graph theory
following~\cite{FB-JC-SM:09}.  A weighted \emph{graph}, is a triplet  $\GG =
(\VV ,\EE ,\vect{\sf{A}})$, where $\VV=\{1,\dots,N\}$ is 
the \emph{node set},
$\EE \subseteq \VV\times \VV$ is the \emph{edge set}, and $\vect{\sf{A}}=[\mathsf{a}_{ij}]\in\real^{N\times N}$ is a weighted \emph{adjacency}
matrix such that $ \mathsf{a}_{ij} >0$ if $(i, j) \in\EE$ and $
\mathsf{a}_{ij} = 0$, otherwise. An edge from
$i$ to $j$, denoted by $(i,j)$, means that agent $j$ can send
information to agent $i$. A graph is \emph{undirected} if $(i,j) \in \EE$ anytime
$(j,i)\in\EE$.  
  An  undirected graph whose weights satisfy $\mathsf{a}_{ij} = \mathsf{a}_{ji}$ for all $i,j\in\VV$ is called a \emph{connected graph} if there is a  path from every node to every other node in the network.
The (out-)\emph{Laplacian} matrix of a graph is $\lL =
\Diag{\vect{\mathsf{A}} \vect{1}_N}- \vect{\mathsf{A}}$.
Note that $\lL\vect{1}_N=\vect{0}$.  
 A graph is connected
 if and only if $\vect{1}_N^\top\lL=\vect{0}$,  and $\rank(\lL)=N-1$. 
 Therefore, for a connected graph zero is a simple eigenvalue of 
$\lL$. 
For a connected graph, we denote the eigenvalues of $\lL$ by $\lambda_1, \dots,\lambda_N$, where $\lambda_1=0$ and $\lambda_i\leq \lambda_j$, for $i<j$. 

\section{Distributed Continuous-Time 
Solvers}\label{sec::Alg}
\vspace{-0.13in}
In this section, we present our distributed algorithm to first solve the constrained optimization problem~\eqref{eq::prob_def} when there is no inequality constraint, i.e., $\underline{\mathcal{B}}^i=\bar{\mathcal{B}}^i=\{\}$ for $i\in\VV$. Then, we extend our results to solve the constrained optimization problem~\eqref{eq::prob_def} with inequality constraints. Our standing assumptions are given below.
\begin{assump}\longthmtitle{Problem specifications}\label{eq::assump_main-1} {\rm
    The  cost function  $f_l^i:\real\to\real$ of the subagent $l\in\mathbb{Z}_{1}^{n^i}$ of each agent $i\in\VV$ is convex and differentiable. Moreover, $\nabla f^i:\real^{n^i}\to\real^{n^i}$ of each agent $i\in\VV$ is locally Lipschitz. Also,
    \begin{align}\label{eq::W_def}
    \vectsf{W}=[\vectsf{w}^1,\dots, \vectsf{w}^N]\in\real^{p\times m}    
    \end{align} 
    is full row rank and the feasible set
 \begin{align}\label{eq::feas}
     \vect{X}_{\text{fe}}=\left\{\vect{x}\in\real^m \,|\, \eqref{eq::prob_def-equal},\eqref{eq::prob_def-box1},\eqref{eq::prob_def-box2}
 \text{~hold}\,\right\}
 \end{align}
 is non-empty for  local inequalities~\eqref{eq::prob_def-box1}
 and~\eqref{eq::prob_def-box2}. Lastly, the optimization problem~\eqref{eq::prob_def} has a finite optimum $f^\star=f(\vectsf{x}^\star)=\sum\nolimits_{i=1}^N f^i(\vectsf{x}^{i\star})$.}\boxend
 \end{assump}
Local Lipschitzness of $\nabla f^i$,  $i \in\mathcal{V}$, guarantees existence and uniqueness of the solution of our proposed  algorithm~\eqref{eq::Alg}, which is a differential equation.

\vspace{-0.05in}
To solve  problem~\eqref{eq::prob_def} subject to only the equality constraints, we consider the augmented cost function with a penalty term on violating the affine constraint,~i.e., \begin{subequations}\label{eq::prob_def-aug2} 
\begin{align}
 \xs&=\underset{\vect{x}\in
    \reals^m}{\argmin}\sum\nolimits_{i=1}^N f^i(\vect{x}^i)+\frac{\rho}{2}\,\|\vectsf{W}\vect{x}-\vect{b}\|^2\!,\label{eq::prob_def-aug2-cost1}\\
&~~~[\vectsf{w}^1]_k\vect{x}^1 +\cdots+[\vectsf{w}^N]_k\vect{x}^N=\bs_k,\quad k\in\mathbb{Z}_{1}^p,\label{eq::prob_def-aug2-constraint} 
     \end{align}
\end{subequations}
where $\rho\in\real_{\geq0}$ is the penalty parameter.
This augmentation results in the so-called  \emph{augmented Lagrangian} formulation of iterative optimization algorithms. As stated in~\cite{SB-NP-EC-BP-JE:10}, augmented Lagrangian methods were developed in part to bring robustness to the dual ascent method, and in particular, to yield convergence
without assumptions like strict convexity or finiteness of the cost function (see also~\cite{DPB-JNT:97}). As shown below, such positive effects are valid also for the continuous-time algorithms we study. Augmenting the cost with the  penalty function as in~\eqref{eq::prob_def-aug2-cost1} however presents a challenge in design of distributed solutions as the total cost in~\eqref{eq::prob_def-aug2-cost1} is no longer separable. Nevertheless, we are able to address this challenge in our distributed solution. 

\begin{lem}\longthmtitle{KKT conditions to characterize solution set of~\eqref{eq::prob_def-aug2}~\cite{SB-LV:04}}\label{lem::sol-op}
{\rm Consider the constrained optimization problem~\eqref{eq::prob_def-aug2}. Let Assumption~\ref{eq::assump_main-1} hold and  $f^i:\mathbb{R}^{n^i}\to\mathbb{R}$, $i\in\VV$, be a differentiable and convex function on $\real^{n^i}$. For any $\rho\in\realnonnegative$, a point $\xs\in\real^m$ is a solution of~\eqref{eq::prob_def-aug2} if and only if there exists a $\nus\in\real^p$, such that, for $i\in\VV$,
\begin{subequations}\label{eq::KKT_1}
\begin{align}
&\nabla f^i(\vectsf{x}^{i\star})+ \vectsf{w}^{i\top}\!\nus=\vect{0},\\
&[\vectsf{w}^1]_k\vectsf{x}^{1\star}+\cdots+[\vectsf{w}^N]_k\vectsf{x}^{N\star}=\bs_k,\quad k\in\mathbb{Z}_{1}^p.
 \end{align}
 \end{subequations}
Moreover, $\nus$ corresponding to every $\xs$ is unique and finite. If the local cost functions are strongly convex, then for any $\rho\in\realnonnegative$ the KKT equation~\eqref{eq::KKT_1} has a unique solution $(\nus,\xs)$, i.e.,~\eqref{eq::prob_def-aug2} has a unique solution. }\boxend
\end{lem}

 Let $L(\vect{\nu},\vect{x})\!=\!  f(\vect{x})+\frac{\rho}{2} \|\vectsf{w}^1\vect{x}^1 \!+\!\cdots\!+\!\vectsf{w}^N\vect{x}^N\!\!-\!\vect{\bs}\|^2+\vect{\nu}^\top(\vectsf{w}^1\vect{x}^1 \!+\!\cdots\!+\!\vectsf{w}^N\vect{x}^N\!\!-\!\vect{\bs})$
be the augmented Lagrangian of the optimization problem~\eqref{eq::prob_def-aug2}. Following~\cite{KJA-LH-HU:58},   
a central solver for the optimal resource allocation problem~\eqref{eq::prob_def-aug2} is
\begin{subequations}\label{eq::saddle-aug}
\begin{align}
\dot{\nu}_k&=\frac{\partial L(\vect{\nu},\vect{x})}{\partial {\nu}_k}=[\vectsf{w}^1]_k\vect{x}^1 \!\!+\!\cdots\!+\![\vectsf{w}^N]_k\vect{x}^N\!\!-\!\mathsf{b}_k,\label{eq::saddle-aug-Lag}\\
\dvect{x}^i&=-\frac{\partial L(\vect{\nu},\vect{x})}{\partial \vect{x}^i}=-\nabla f^i(\vect{x}^i)-\sum\nolimits_{j=1}^p[\vectsf{w}^{i}]_j^{\top}\,\nu_j\,-\label{eq::saddle-aug-x}\nonumber\\
&~\qquad\quad\qquad\qquad \rho\,{\vect{w}^{i\top}}(\vectsf{w}^1\vect{x}^1 \!\!+\!\cdots\!+\!\vectsf{w}^N\vect{x}^N\!\!-\!\vect{\bs}),
\end{align}
\end{subequations}
where $k\in\mathbb{Z}_1^p$, and $i\in\VV$. The  algorithm studied in~\cite{KJA-LH-HU:58} is for un-augmented Lagrangian, i.e., $\rho=0$, and the guaranteed convergence holds only for strictly convex cost function $f(\vect{x})$. However, we can show that the central solver~\eqref{eq::saddle-aug} with $\rho>0$ is guaranteed to converge for convex cost function $f(\vect{x})$, as well (the details are omitted for brevity). A numerical example demonstrating this positive role is presented in Appendix~B.

The source of coupling in~\eqref{eq::prob_def-aug2} is the set of the equality constraints~\eqref{eq::prob_def-aug2-constraint}, which appear in the central solver~\eqref{eq::saddle-aug}, as well. To design our distributed algorithm, we adapt the structural constitution of~\eqref{eq::saddle-aug}, but aim to create the coupling terms $[\vectsf{w}^1]_k\vect{x}^1 \!\!+\!\cdots\!+\![\vectsf{w}^N]_k\vect{x}^N\!-\mathsf{b}_k$, $k\in\mathbb{Z}_1^p$, in a distributed manner. We note that for every equality constraint $k\in\mathbb{Z}_1^p$, the coupling is among the set of agents $\mathcal{C}_k=\{i\in\VV\,|\, [\vectsf{w}^i]_k\neq\vect{0}\}$. To have an efficient communication and computation resource management, we seek an algorithm that handles every coupled equality constraint among only those agents that are involved. In this regards, for every equality constraint $k\in\mathbb{Z}_1^p$, we let $\mathcal{G}_k(\VV_k,\EE_k)$ be a connected undirected subgraph of $\GG$ that contains the set of agents $\mathcal{C}_k$ (see Fig.~\ref{fig::network} for an example). We assume that $\VV_k\subset\VV$ is a monotonically increasing ordered set. It is very likely that the agents coupled through an equality constraint are geographically close, and thus in the communication range of each other. Nevertheless, $\VV_k$, $k\in\mathbb{Z}_1^p$, may contain agents $i\in\VV$ that have $[\vectsf{w}^i]_k=\vect{0}$ but are needed to make $\GG_k$ connected (see Fig.~\ref{fig::network} for an example). 
We let $N_k=|\VV_k|$, $k\in\mathbb{Z}_{1}^p$. 
In our distributed solution for~\eqref{eq::prob_def-aug2}, we also seek an algorithm that allows each agent to use a local penalty parameter $\rho^i\in\real_{>0}$, so we can eliminate the need to coordinate among the agents to choose the penalty parameter $\rho$. In what follows, we define  $\mathcal{T}^i=\{j\in\mathbb{Z}_1^p|i\in\VV_j\}$, $i\in\VV$, and $\{\bar{\mathsf{b}}_k^l\}_{l\in\VV_k}$ such that  $\sum_{l\in\VV_k}{\bar{\mathsf{b}}}_k^l=\bs_k$, for $k\in\mathbb{Z}_1^p$ (possible options include   $\bar{\mathsf{b}}_k^l=\bs_k/|\mathcal{C}_k|$, $l\in\mathcal{C}_k$ while $\bar{\mathsf{b}}_k^j=0$, $j\in\VV\backslash\mathcal{C}_k$, or   $\bar{\mathsf{b}}^j_k=\bs_k$ for a particular agent $j\in\VV_k$ and $\bar{\mathsf{b}}^l_k=0$ for any $l\in \VV\backslash\{j\}$).

With the right notation at hand, our  proposed distributed algorithm to solve  optimization problem~\eqref{eq::prob_def-aug2}~is
\begin{subequations}\label{eq::Alg}
  \begin{align}
    \dot{y}^l_k =\, &\beta_k\sum\nolimits_{j\in\VV_k}
    \mathsf{a}_{l j} (v^l_k-v_k^j), \label{eq::Alg-y}
    \\
    \dot{v}^l_k =\, &(
    [\vectsf{w}^l]_{k}    \vect{x}^l-\bar{\bs}^l_k)\!-\!\beta_k\!\,\sum\nolimits_{j\in\VV_k}
    \!\!\mathsf{a}_{l j}(v^l_k-v^j_k)
    -y_k^l,\label{eq::Alg-z}\\
     \dot{\vect{x}}^i=&-(1+\rho^i)\nabla f^i(\vect{x}^i) \!-\!\rho^i\sum\nolimits_{k\in \mathcal{T}^i}\!{[\vectsf{w}^i]_k^\top}([\vectsf{w}^i]_{k}\vect{x}^i\!-\!\bar{\bs}^i_k)\nonumber\\ +\rho^i\,&\sum\limits_{k\in \mathcal{T}^i}{([\vectsf{w}^i]^\top_{k}}y^i_k)
     \!-\!(1+\rho^i)\!\,\sum\limits_{k\in \mathcal{T}^i}({[\vectsf{w}^i]^\top_{k}}v^i_k),\label{eq::Alg-x}
  \end{align}
\end{subequations}
with $\beta_k\!\in\!\real_{>0}$ and $\rho^i\!\in\!\real_{\geq0}$ for $i\!\in\!\VV$, $k\in\mathbb{Z}_1^p$ and $l\in\VV_k$. To comprehend the connection with the centralized dynamical solver~\eqref{eq::saddle-aug}, take summation of~\eqref{eq::Alg-y} and~\eqref{eq::Alg-z} over every connected  $\GG_k,\,k\in\mathbb{Z}_1^p$ to obtain
\begin{subequations}\label{eq::dis_sol_mul}
\begin{align}
  &\sum\nolimits_{l\in\VV_k}\!\!\dot{y}^l_k=0 \Longrightarrow\label{eq::dis_sol_mul_a}
  \sum\nolimits_{l\in\VV_k}\!\!{y}^l_k(t)=\sum\nolimits_{l\in\VV_k}\!\!{y}^l_k(0),\\
  &\sum\nolimits_{l\in\VV_k}\!\!\dot{v}^l_k=[\vectsf{w}^1]_k\vect{x}^1 \!\!+\!\cdots\!+\![\vectsf{w}^N]_k\vect{x}^N\!\!-\!\mathsf{b}_k,
\end{align}
\end{subequations}
which shows that for any  $k\in\mathbb{Z}_1^p$, the dynamics of the sum of  $v^l_k$s duplicates the Lagrange multiplier dynamics~\eqref{eq::saddle-aug-Lag} of the central
Augmented Lagrangian method. Therefore, in a convergent~\eqref{eq::Alg},
ultimately for each $k\in\mathbb{Z}^{p}_1$, all the $v^l_k$s 
converge to the same value indicating
that ultimately every agent obtains a  local copy
of~\eqref{eq::saddle-aug-Lag} for any $k\in\mathbb{Z}^{p}_1$. On the other hand, if we factor out $(1+\rho^i)$ from the right hand side of~\eqref{eq::Alg-x} and exclude the third component, which is a technical term added to induce agreement between the agents,~\eqref{eq::Alg-x} mimics the dynamics~\eqref{eq::saddle-aug-x} of the central Augmented Lagrangian~solver.


  \begin{rem}(Benefits of cluster-based approach)
\rm{First we note that regardless of the size of $n^i$, in algorithm~\eqref{eq::Alg} we associate at most 
one copy of the Lagrange multiplier generator dynamics, i.e.,~\eqref{eq::Alg-y} and~\eqref{eq::Alg-z}, to every agent $i\in\VV$. Specifically, every agent $i\in\VV$, maintains $|\mathcal{T}^i|\leq p$ number of~\eqref{eq::Alg-y} and~\eqref{eq::Alg-z} pair dynamics and consequently has to broadcast the same number of variables to the network. In comparison, if we use the algorithms in~\cite{LX-SB:06,YZ-MMZ:18,SAA-KY-AHS:18,PY-YH-FL:16,SSK:17,DD-MRJ:18,AC-JC:16-auto,AC-JC:15-tcns}, when $n^i>1$, for any $i\in\mathcal{V}$, we need to treat each component of the $i$ as an agent and assign a copy of a dynamics that generates the dual variable to every subagent $l\in\mathbb{Z}_1^{n^i}$. This results in a storage, computation and communication cost of order $n^i\times p$ per agent $i\in\mathcal{V}$. See our numerical examples for a comparison. Next, notice that algorithm~\eqref{eq::Alg} can always be implemented by using $\mathcal{G}_k=\mathcal{G}$, $k\in\mathbb{Z}_1^p$, where $\mathcal{G}=(\mathcal{V},\mathcal{E})$ is the connected interaction topology that all the agents form. However, the flexibility to use a smaller cyber-layer formed by only the cluster of agents that are coupled by an equality constraint reduces the communication and computational cost of implementing Algorithm~\eqref{eq::Alg}. Moreover, in some problems, similar to our numerical example in Section~\ref{sec::num}, the coupling equation is between the neighboring agents. In such cases, subgraphs $\mathcal{G}_k$ can be easily formed. Moreover, as one can expect and our  numerical example also highlights, using a smaller subgraph $\mathcal{G}_k$ can results in a faster convergence for~\eqref{eq::Alg-y} and ~\eqref{eq::Alg-z} dynamics and as a result a faster convergence for algorithm~\eqref{eq::Alg}. \boxend
}
\end{rem}

\vspace{-0.05in}
The equilibrium points of algorithm~\eqref{eq::Alg} when every $\GG_k$, $k\in\mathbb{Z}_{1}^p$ is a connected graph is given by 
\begin{align}\label{eq::eqillibria}
\!\!\!\mathcal{S}_e\!=\!\Big\{&(\{\!\vect{v}_k\}_{k=1}^p,\{\vect{y}_k\}_{k=1}^p,\{\vect{x}^i\}_{i=1}^N)\in\!\prod_{k=1}^p\real^{N_k}\!\times\prod_{k=1}^p\real^{N_k}\!\times\nonumber\\&\!\!\!\!\!\!\!\!\!\!\!\!\!\!\prod_{i=1}^N\real^{n^i}\Big|\vect{v}_k=
\theta_k\vect{1}_{N_k},\theta_k\in\real,\, \nabla f^i(\vect{x}^i)\!+\!\!\!\sum_{j\in\mathcal{T}^i}[\vectsf{w}^i]^\top_j\theta_j\!=\!\vect{0},\nonumber\\&\!\!\!\!\!\!\!\!\!\!\!\! \sum\nolimits_{j=1}^N[\vectsf{w}^j]_k \vect{x}^j\!=\!\bs_k+\!\!\sum\nolimits_{j\in\VV_k}y_k^j,~y_k^l=[\vectsf{w}^l]_k \vect{x}^l-\bar{\bs}^l_k,\nonumber\\&~~\,\,\quad\qquad\qquad\qquad\quad ~i\in\VV,l\in\VV_k, k\in\mathbb{Z}_1^p
\Big\}.
\end{align}
Due to~\eqref{eq::dis_sol_mul_a}, if algorithm~\eqref{eq::Alg} is initialized such that $\sum\nolimits_{l\in\VV_k}y_k^l(0)=0$, we have 
$\sum\nolimits_{l\in\VV_k}{y}^l_k(t)=\sum\nolimits_{l\in\VV_k}{y}^l_k(0)$ for $t\in\real_{\geq0}$.
In that case, if algorithm~\eqref{eq::Alg}  
converges to an equilibrium point
$(\{\Bvect{v}_k\}_{k=1}^p,\{\bar{\vect{y}}_k\}_{k=1}^p, \{\Bvect{x}^i\}_{i=1}^N)\in\mathcal{S}_e$, 
we have
$(\{\Bvect{v}_k\}_{k=1}^p, \{\bar{\vect{y}}_k\}_{k=1}^p, \{\Bvect{x}^i\}_{i=1}^N)\!=\!(\{[\{[\vectsf{w}^l]_k \vectsf{x}^{l\star}-\bar{\bs}^l_k\}_{l\in\VV_k}]\}_{k=1}^p,\,\{\nu^\star_k \vect{1}_{N_k}\}_{k=1}^p,\{\vectsf{x}^{i\star}\}_{i=1}^N)$, where $(\{\vectsf{x}^{i\star}\}_{i=1}^N,\\\{\nu^\star_k\}_{k=1}^p)$ satisfies the KKT equation~\eqref{eq::KKT_1}. The following theorem shows that indeed under the stated initialization, the algorithm~\eqref{eq::Alg} converges to a minimizer of optimization problem~\eqref{eq::prob_def-aug2}. 
To establish the proof of this theorem we use the following notations. We let $\vectsf{A}\in\real^{N\times N}$ be the adjacency matrix of $\GG$. Then, the the adjacency matrix of $\GG_k\subset\GG$, $k\in\mathbb{Z}_{1}^p$, is $\vectsf{A}_k$, which is the submatrix of $\vectsf{A}$ corresponding to the rows and the columns associated with the agents in $\VV_k$, i.e., $\vectsf{A}_k=\vect{M}_k^\top\,\vectsf{A}\,\vect{M}_k$ where $\vect{M}_k\in\real^{N\times N_k}$ is defined such that $[\vect{M}_k]^l=[\vectsf{I}]^{\VV_k(l)}$, $l\until{N_k}$ with $\VV_k(l)$ being the $l^{th}$ element of the ordered set $\VV_k$. Then, $\lL_k=\Diag{\vectsf{A}_k \vect{1}_{N_k}}- \vectsf{A}_k$ is the Laplacian matrix of $\GG_k$, $k\in\mathbb{Z}_1^p$. 
Next, we define 
  $\rr_k=\frac{1}{\sqrt{N_k}}\vect{1}_{N_k}$ and $\rR_k=[\vect{v}_{2k},\cdots,\vect{v}_{N_kk}]$ with $(\rr_k, \{\vect{v}_{jk}\}_{j=2}^{N_k})$ being the normalized eigenvectors of $\lL_k$. Note here that we have
  \begin{subequations}
  \begin{align}
  &\rr_k^\top\rR_k \!=\!\vect{0},~\rR_k^\top\rR_k=
  \iI_{N_k-1},~ \rR_k\rR_k^\top\!
  =\!\pPi_{N_k},~\\
  &[\vect{\mathsf{r}}_k\,\, \rR_k]^\top\lL_k[\vect{\mathsf{r}}_k\,\, \rR_k]=\Diag{[0,\lambda_{2k},\cdots,\lambda_{N_kk}]}.
  \end{align}
  \end{subequations}
The eigenvectors are ordered such that $\lambda_{2k}$ and $\lambda_{N_kk}$ are, respectively, the smallest and the largest non-zero eigenvalues of $\lL_k$. The next two theorems whose proofs are given in~Appendix~A examine the stability and convergence of~\eqref{eq::Alg} over connected graphs.

 \begin{thm}\longthmtitle{Asymptotic convergence of~\eqref{eq::Alg} over connected graphs when the local costs are convex}\label{thm::main}
  {\rm Let every $\GG_k$, $k\in\mathbb{Z}_1^p$, be a connected graph and Assumption~\ref{eq::assump_main-1} hold. For every $k\in\mathbb{Z}_1^p$, suppose $\{\bar{\mathsf{b}}_k^l\}_{l\in\VV_k}\subset\real$ is defined such that  $\sum_{l\in\VV_k}{\bar{\mathsf{b}}}_k^l=\bs_k$. Then, for each $i\in\VV$, $l\in\VV_k$, starting from  $\vect{x}^i(0)\in\real^{n^i}$ and $y^l_k(0),v^l_k(0)\in\real$   with
  $\sum_{l\in\VV_k}y^l_k(0) \!=\! 0$, the
  algorithm~\eqref{eq::Alg} for any $\rho^i\in\realpositive$, makes
  $t\mapsto(\{\vect{v}_k(t)\}_{k=1}^p,\{\vect{x}^i(t)\}_{i=1}^N)$ converge asymptotically to $(\,\{{{\nu}}_k^\star\vect{1}_{N_k}\}_{k=1}^p,\{{\vectsf{x}^{i\star}}\}_{i=1}^N)$, where $(\{{\nu}_k^\star\}_{k=1}^p,\{{\vectsf{x}^{i\star}}\}_{i=1}^N)$ is a point satisfying the KKT  conditions~\eqref{eq::KKT_1}  of problem~\eqref{eq::prob_def-aug2}.} \boxend
   \end{thm}
The initialization condition $\sum_{l\in\VV_k}y^l_k(0)=0$ of Theorem~\ref{thm::main} is trivially satisfied by every agent $l\in\VV_k$, $k\in\mathbb{Z}_1^p$, using $y^l_k(0)=0$.  The asymptotic convergence guarantee for  algorithm~\eqref{eq::Alg} in Theorem~\ref{thm::main} is established for local convex cost functions. For such cost functions, similar to the centralized algorithm~\eqref{eq::saddle-aug},~\eqref{eq::Alg} fails to converge when $\rho^i=0$ for all $i\in\VV$. Next, we show that if the local costs are strongly convex and have Lipschitz gradients then the convergence is in fact exponentially fast for $\rho^i\in\real_{>0}$ $i\in\VV$.  Recall that for strongly convex local cost functions, the minimizer of~\eqref{eq::prob_def-aug2}  is unique.
 \begin{thm}\longthmtitle{Exponential convergence of~\eqref{eq::Alg} over connected graphs when the local costs are strongly convex and have Lipschitz gradients }\label{prop::main-exp}
 {\rm Let every $\GG_k$, $k\!\in\!\mathbb{Z}_1^p$  be  connected and Assumption~\ref{eq::assump_main-1} hold. Also, assume each cost function $f^i_l$, $l\!\in\!\mathbb{Z}_1^{n^i}$, $i\!\in\!\VV$, is $m^i_l$-strongly convex and has $M^i_l$-Lipschitz gradient. Let  $m\!=\!\max\{\{m^i_l\}_{l=1}^{n^i}\}_{i=1}^{N}\in\real_{>0}$ and $M\!=\!\max\{\{M^i_l\}_{l=1}^{n^i}\}_{i=1}^N\in\realpositive$. Then,  starting from  $\vect{x}^i(0)\!\in\!\real^{n^i}$ and $y_k^l(0),v^l_k(0)\!\in\!\real$ for each $i\!\in\!\VV$, $l\in\VV_k$,  and given
  $\sum_{l\in\VV_k}y^l_k(0)\!=\!0$ and $\sum\nolimits_{l\in\VV_k}\bar{b}_k^l\!=\!\bs_k$ in~\eqref{eq::Alg}, the
  algorithm~\eqref{eq::Alg} makes
  $t\mapsto(\{\vect{v}_k(t)\}_{k=1}^p,\{\vect{x}^i(t)\}_{i=1}^N)$ converge exponentially fast to $(\,\{{{\nu}}_k^\star\vect{1}_{N_k}\}_{k=1}^p,\{{\vectsf{x}^{i\star}}\}_{i=1}^N)$  for any $\rho^i\in\real_{>0}$, where $(\{{\nu}_k^\star\}_{k=1}^p,\{{\vectsf{x}^{i\star}}\}_{i=1}^N)$ is the unique solution of the KKT conditions~\eqref{eq::KKT_1} of problem~\eqref{eq::prob_def-aug2}. Moreover, when $\rho^i=0$ for an  $i\in\VV$, the convergence to the unique solution of the KKT conditions~\eqref{eq::KKT_1} is~asymptotic.}\boxend
   \end{thm}
 The proof of Theorem~\ref{prop::main-exp} is given in Appendix~A.

\begin{rem}(The convergence of~\eqref{eq::Alg} over dynamically changing connected graphs)
\rm{The proof of Theorem~\ref{prop::main-exp} relies on a Lyapunov function that is independent of the systems parameters, and its derivative for $\rho^i\in\real_{>0}$, $i\in\VV$, is negative definite with a quadratic upper bound. Hence, we can also show that the algorithm~\eqref{eq::Alg}, when $\rho^i\in\real_{>0}$ for $i\in\VV$, converges exponentially fast to a unique solution of the KKT conditions~\eqref{eq::KKT_1} of problem~\eqref{eq::prob_def-aug2} over any time-varying topology $\GG_k$, $k\in\mathbb{Z}_1^p$ that is connected at all times and its adjacency matrix is uniformly bounded and piece-wise constant.}
\end{rem}

\subsection{Problem subject to both equality and inequality constraints}
\vspace{-0.1in}
To address inequality constraints, 
we use a penalty function method to eliminate the local inequality constraints~\eqref{eq::prob_def-box1} and~\eqref{eq::prob_def-box2}. 
That is, we seek solving 
\begin{subequations} \label{eq::prob_def_penala}
\begin{align}
  \vectsf{x}_p^\star=&\arg\min_{\vect{x}\in\real^m} \,\,\sum\nolimits_{i=1}^N f_p^i(\vect{x}^i),~~\text{subject~to~} \\
    &~ [\vectsf{w}^1]_j\vect{x}^1+\cdots+[\vectsf{w}^N]_j\vect{x}^N=\bs_j,\quad
    j\in\mathbb{Z}_{1}^{p},\label{eq::prob_def_penalb} 
\end{align}
\end{subequations}
with
\begin{align}\label{eq::smooth_penalty}\!\!\!f^i_{\text{p}}(\vect{x}^i)\!=\!\!f^i(\vect{x}^i)\!+\!\gamma\big(\!\sum_{l\in\underline{\mathcal{B}}^i}\! p_\eps(\bunderline{\mathsf{x}}^i_l\!-\!x^i_l)\!+\!\!\sum_{l\in\bar{\mathcal{B}}^i} \!p_\eps(x^i_l\!-\!\bar{\mathsf{x}}^i_l)\big),\end{align} 
$i\in\VV$, where $\gamma\in\real_{>0}$ is the weight of the smooth penalty function $p_\eps=\begin{cases}0,& y\leq 0,\\
\,\frac{1}{2\eps}y^2,&0\leq y\leq \eps,\\
(y-\frac{1}{2}\eps),& y\geq \eps,
\end{cases}$ 
for some $\eps\in\!\realpositive$. This approach allows us to use algorithm~\eqref{eq::Alg} to solve the optimization~\eqref{eq::prob_def} by using $f^i_{\text{p}}(\vect{x}^i)$ in place of $f^i(\vect{x}^i)$ in~\eqref{eq::Alg-x}. We note that $f^i_{\text{p}}(\vect{x}^i)$ is convex and differentiable if $f^i(\vect{x}^i)$ is a convex function in $\real^{n^i}$.  Following this penalty method approach, when the global cost function of~\eqref{eq::prob_def} is evaluated at the limit point of  algorithm~\eqref{eq::Alg}, it is in $\epsilon$-order neighborhood of the global optimal value of the optimization problem~\eqref{eq::prob_def} (see Proposition~\ref{prop::penalty} below). In what follows, we investigate when the penalty function weight $\gamma$ has a finite value and give a well-defined admissible range for it. 

Given Assumption~\ref{eq::assump_main-1}, the Slater condition~\cite{SB-LV:04} is satisfied. Thus,
the KKT conditions below give~a set of necessary and sufficient conditions that~characterize the solution set of the convex optimization problem~\eqref{eq::prob_def}. 

\begin{lem}\longthmtitle{Solution set of~\eqref{eq::prob_def}~\cite{SB-LV:04}}\label{lem::sol-op}
{\rm Consider the constrained optimization problem~\eqref{eq::prob_def} under Assumptions~\ref{eq::assump_main-1}. 
A point $\xs\in\real^m$ is a solution of~\eqref{eq::prob_def} if and only if there exists $\nus\in\real^p$ and $\{\bunderline{{\mu}}_l^{i\star}\}_{l\in\underline{\mathcal{B}}^i}\subset\real_{\geq0}$
$\{\bar{{\mu}}_l^{i\star}\}_{l\in\bar{\mathcal{B}}^i}\subset\real_{\geq0}$, $i\in\VV$, such that 
\begin{subequations}\label{eq::KKT1}
\begin{align}
&\nabla f^i(\xsi)\!+\!\vectsf{w}^{i\top}\nus-\bunderline{\vect{\mu}}^{i\star}+\bar{\vect{\mu}}^{i\star}=\vect{0},\label{eq::KKT_a}\\
&\vectsf{W}\xs-\vectsf{b}=\vect{0}, \\
&\bunderline{\mu}^{i\star}_l(\bunderline{\mathsf{x}}_l^i\!-\!\mathsf{x}_l^{i\star})\!=\!0,~\bunderline{\mathsf{x}}^i_l\!-\!\mathsf{x}^{i\star}_l\leq\! 0,~ \bunderline{\mu}_l^{i\star}\!\geq\!0,~~l\!\in\!\underline{\mathcal{B}}^i,\label{eq::KKT-box1}\\
&\bar{\mu}_{l}^{i\star}(\mathsf{x}_l^{i\star}\!-\!\bar{\mathsf{x}}_l^i)\!=\!0,~\mathsf{x}^{i\star}_l\!-\! \bar{\mathsf{x}}^i_l\leq\!0,~ \bar{\mu}_{l}^{i\star}\!\geq\!0,\,\, ~l\!\in\!\bar{\mathcal{B}}^i,\label{eq::KKT-box2}
 \end{align}
 \end{subequations}
 where $\bunderline{\vect{\mu}}^{i\star}=[\bunderline{\mu}_1^{i\star},\cdots,\bunderline{\mu}_{n^i}^{i\star}]^\top$ with $\bunderline{\mu}_l^{i\star}=0$ for $l\in\mathbb{Z}_{1}^{n^i}\backslash\underline{\mathcal{B}}^i$ and $\bar{\vect{\mu}}^{i\star}=[\bar{\mu}_1^{i\star},\cdots,\bar{\mu}_{n^i}^{i\star}]^\top$ with $\bar{\mu}_l^{i\star}=0$ for $l\in\mathbb{Z}_{1}^{n^i}\backslash\bar{\mathcal{B}}^i$. 
If the local cost functions are strongly convex, then the optimization problem~\eqref{eq::prob_def} has a unique solution.
  \boxend}
\end{lem}

Let $X^\eps_{\text{fe}}$ be the $\eps$-feasible set of optimization problem~\eqref{eq::prob_def},
\begin{align}\label{eq::eps-feasible}
X^\eps_{\text{fe}}=\big\{\vect{x}\in\real^m\,|\,& \vectsf{W}\vect{x}=\vectsf{b},~~\bunderline{\mathsf{x}}^i_l\!-\!{x}^i_l\leq\! \eps,~l\!\in\!\underline{\mathcal{B}}^i\nonumber\\
&~ \qquad{x}^i_j\!-\! \bar{\mathsf{x}}^i_j\leq\!\eps,~j\!\in\!\bar{\mathcal{B}}^i,~ i\in\VV\big\}.
\end{align}
The result below states that for some admissible values of $\gamma$, the minimizer of problem~\eqref{eq::prob_def_penala} belongs to $\eps$-feasible set $X^\eps_{\text{fe}}$ and optimal value of optimization problem~\eqref{eq::prob_def} is in $\eps$ order neighborhood of the optimal value of the original optimization problem~\eqref{eq::prob_def}.
\begin{prop}\longthmtitle{relationship between the  solution of~\eqref{eq::prob_def} and~\eqref{eq::prob_def_penala}~\cite{MCP-SAZ:94}}\label{prop::penalty}
{\rm Let $(\xs,\nus,\{\bunderline{{\mu}}_l^{i\star}\}_{l\in\underline{\mathcal{B}}^i},\{\bar{{\mu}}_l^{i\star}\}_{l\in\bar{\mathcal{B}}^i})$ be any solution of the KKT equations~\eqref{eq::KKT_1}. Let ${\vectsf{x}}_{\text{p}}^\star$ be a minimizer of optimization problem~\eqref{eq::prob_def_penala} for some $\gamma,\eps\in\realpositive$. If $\gamma=\frac{1-N}{1-\sqrt{N}}\gamma^\star$, where $\gamma^\star> \max\big\{\max\{\bunderline{\mu}_l^{i\star }\}_{l\in\underline{\mathcal{B}}^i},\max\{\bar{\mu}_l^{i\star }\}_{l\in\bar{\mathcal{B}}^i}\big\}_{i=1}^{N}$, 
then 
\begin{align}\label{eq::eps-feas-guarantee}
&{\vectsf{x}}_{\text{p}}^\star\in  X^\eps_{\text{fe}},\quad 0\leq f^\star-f({\vectsf{x}}_{\text{p}}^\star)\leq \eps\,\gamma N,
\end{align}
where $f^\star=f(\xs)$ is the optimal value of~\eqref{eq::prob_def}.}
\boxend
\end{prop}
We note that if $\eps\!\to\!0$, we have $p_\eps(y)\!\to\! p(y)=\max\{0,y\}$, where $p(y)$~is the well-known non-smooth penalty function~\cite{DPB:75} with exact equivalency guarantees  when $\gamma\!>\!\gamma^\star$ in Proposition~\ref{prop::penalty}.

\begin{rem}(comment on the feasibility of solution of~\eqref{eq::prob_def_penala})\label{rem::adjusted}
{\rm Use of $\eps-$exact penalty function approach is motivated by keeping the cost  smooth and differentiable, which is of desire from practical perspective compared to exact penalty method which is a non-smooth function. Using an $\eps$-exact penalty function we have the grantees that the approximated solution $\vectsf{x}^{\star}_p$ is in~\eqref{eq::eps-feasible}. Therefore only the inequality constrains may be violated by $\eps$ amount. Since the value of $\eps$ can be selected very small, the possible violation of the inequality constraints will be  small too. One may select the value of $\eps$ in accordance to the expected accuracy of the algorithm. 
Note that by slight tightening of the inequality constraints according to
${x}^i_l\leq \bar{\mathsf{x}}^i_l-\eps$ and $\bunderline{\mathsf{x}}^i_l+\eps\leq{x}^i_l$ and using these adjusted inequalities in the penalty function, we can guarantee that ${\vectsf{x}}_{\text{p}}^\star\in  X_{\text{fe}}$. But this may result in slight increase in the optimally gap in~\eqref{eq::eps-feas-guarantee}. 
}
\end{rem}


 Considering Proposition~\ref{prop::penalty}, a practical and numerically well-posed solution via the penalty optimization method~\eqref{eq::prob_def_penala} is achieved when the Lagrange multipliers are bounded.
Thus, in what follows we seek for ${\mu}_{\text{bound}}$ in 
  \begin{align}\label{eq::mu_bound}
      \max\!\big\{\!\max\{\bunderline{\mu}_l^{i\star }\}_{l\in\underline{\mathcal{B}}^i},\max\{\bar{\mu}_l^{i\star }\}_{l\in\bar{\mathcal{B}}^i}\big\}_{i=1}^{N}\leq {\mu}_{\text{bound}},
  \end{align}
with the objective of choosing a penalty function weight $\gamma$ that satisfies the condition set by Proposition~\ref{prop::penalty} by setting $\gamma\geq \frac{1-N}{1-\sqrt{N}}\,{\mu}_{\text{bound}}$.


For any solution of the KKT conditions~\eqref{eq::KKT_1}, we let $\bunderline{\mathcal{A}}^i\subset\bunderline{\mathcal{B}}^i$ and $ \bar{\mathcal{A}^i}\subset\bar{\mathcal{B}^i}$ respectively be the set of indices of the active lower bound and the active upper bound inequality constraints of agent $i\in\VV$. We note that $\bunderline{\mathcal{A}}^i\cap \bar{\mathcal{A}}^i=\{\}$. 
Because for inactive inequalities  
$\bar{\mu}^{i\star}_l=0$ (resp. $\bunderline{\mu}^{i\star}_l=0$) for  $l\in\bar{\mathcal{B}}^i\backslash\bar{\mathcal{A}^i}$ and $i\in\VV$ (resp. $l\in\bunderline{\mathcal{B}}^i\backslash\bunderline{\mathcal{A}}^i$)~\cite{DPB:99}, we obtain
\begin{align}\label{eq::activeset_bound}
&\max\big\{\max\{\bunderline{\mu}_l^{i\star }\}_{l\in\underline{\mathcal{B}}^i},\max\{\bar{\mu}_l^{i\star }\}_{l\in\bar{\mathcal{B}}^i}\big\}_{i=1}^{N}=\nonumber\\
&\quad\quad\quad\max\big\{\max\{\bunderline{\mu}_l^{i\star }\}_{l\in\underline{\mathcal{A}}^i},\max\{\bar{\mu}_l^{i\star }\}_{l\in\bar{\mathcal{A}^i}}\big\}_{i=1}^{N}.
\end{align}
Therefore, to find ${\mu}_{\text{bound}}$,  it suffices to find an upper bound on $\max\big\{\max\{\bunderline{\mu}_l^{i\star }\}_{l\in\underline{\mathcal{A}}^i},\max\{\bar{\mu}_l^{i\star }\}_{l\in\bar{\mathcal{A}}^i}\big\}_{i=1}^{N}$.

As known, the set of the Lagrange multipliers of an optimization problem of form~\eqref{eq::prob_def} is nonempty and bounded if and only if the Mangasarian-Fromovitz constraint qualification (MFCQ) holds~\cite{OLM-SF:67}. It is straight-forward to show that the MFCQ condition is satisfied for a resource allocation problem of form~\eqref{eq::prob_def} with one equality constraint (i.e., $p=1$) and upper and lower bounded decision variables (i.e., $\underline{\mathcal{B}}^i=\bar{\mathcal{B}}^i=\mathbb{Z}_{1}^{n_i}$).
  For such a problem the following result specifies a ${\mu}_{\text{bound}}$ that satisfies~\eqref{eq::mu_bound}.
  
  \begin{prop}\longthmtitle{${\mu}_{\text{bound}}$ for  the resource allocation problem  with one equality constraint and bounded decision variables}\label{prop::mu_bound_weighted_economic_dispach}
  {\rm Consider problem~\eqref{eq::prob_def} under Assumption~\ref{eq::assump_main-1} when $p=1$, $\mathsf{w}^i_l>0$ for $l\in\{1,\cdots,n^i\}$ and $\underline{\mathcal{B}}^i=\bar{\mathcal{B}}^i=\mathbb{Z}_{1}^{n_i}$, $i\in\VV$. Let $(\xs,\nu^\star,\{\bunderline{{\mu}}_l^{i\star}\}_{l\in\underline{\mathcal{B}}^i},\{\bar{{\mu}}_l^{i\star}\}_{l\in\bar{\mathcal{B}}^i})$ be an arbitrary solution of the KKT conditions~\eqref{eq::KKT_1} for this problem. Then, ${\mu}_{\text{bound}}$ in~\eqref{eq::mu_bound} satisfies
\begin{align}\label{eq::mu_bound_p1}
{\mu}_{\text{bound}}\leq\!
(1+\frac{\bar{\mathsf{w}}}{\underline{\mathsf{w}}})\max\big\{\underset{\vect{x}^i\in X^i_{\text{ineq}}}{\max}\,{\|\nabla f^i(\vect{x}^i)\|_\infty}\big\}_{i=1}^N,
\end{align}
where $X^i_\text{ineq}=\{\vect{x}^i\in\real^{n^i}|\, \underline{\mathsf{x}}^i_l\leq{x}^i_l\leq \bar{\mathsf{x}}^i_l,l\in\mathbb{Z}_1^{n^i}\}$,  $\underline{\mathsf{w}}=\min\{\{\mathsf{w}^i_l\}_{l=1}^{n_i}\}_{i=1}^N$ and $\bar{\mathsf{w}}= \max\{\{\mathsf{w}^i_l\}_{l=1}^{n_i}\}_{i=1}^N$.}
\end{prop}
  \begin{pf}
For any given $(\xs,\nu^\star,\{\bunderline{{\mu}}_l^{i\star}\}_{l\in\underline{\mathcal{B}}^i},\{\bar{{\mu}}_l^{i\star}\}_{l\in\bar{\mathcal{B}}^i})$, we note that the KKT conditions~\eqref{eq::KKT_1} can be written as 
\begin{subequations}\label{eq::KKT2_s}
\begin{align}
&\nabla f^i_l(\mathsf{x}^{i\star }_l)\!+\!\mathsf{w}^i_l\,\nu^\star=0,\,\quad\quad l\in\mathbb{Z}_1^{n^i}\backslash\{\bar{\mathcal{A}}^i\cup{\bunderline{\mathcal{A}}^i}\},\label{eq::KKT_a2_s}\\
&\nabla f^i_l(\mathsf{x}^{i\star }_l)\!
+\mathsf{w}^i_l\,\nu^\star+\bar{\mu}^{i\star}_l=0,\,\,\quad\quad l\in\bar{\mathcal{A}}^i,\label{eq::KKT_a3_s}\\
&\nabla f^i_l(\mathsf{x}^{i\star }_l)\!+\mathsf{w}^i_l\,\nu^\star-\bunderline{\mu}^{i\star}_l=0,\,\,~~~\quad l\in\bunderline{\mathcal{A}}^i\label{eq::KKT2_a_s}.
\end{align}
 \end{subequations}
Since $\{\mathsf{w}^i_l\}_{l=1}^{n^i}\subset \real_{>0}$, it follows from Assumption~\ref{eq::assump_main-1}, which states that the feasible set is non-empty for strict local inequalities, that the upper bounds (similarly the lower bounds) for all decision variable  cannot be active simultaneously. Therefore, for any given minimizer, we have either (a) at least for one subagent $k\in\mathbb{Z}_{1}^{n^i}$ in an agent $i\in\VV$ we have $ \underline{\mathsf{x}}^i_k<\mathsf{x}^{i\star}_k< \bar{\mathsf{x}}^i_k$ or (b) some of the decision variables are equal to  their upper bound and the remaining others are equal to their lower bound.
If case (a) holds, it follows from~\eqref{eq::KKT_a2_s} that $\nu^\star=\frac{-\nabla f^i_k(\mathsf{x}^{i\star }_k)}{\mathsf{w}^i_k}$, which means that we have the guarantees that  $|\nu^\star|\leq\frac{\max\{\|\nabla f^i(\vectsf{x}^{i\star})\|_\infty\}_{i=1}^N}{\underline{\mathsf{w}}}$. On the other hand, if (b) holds, then there exists at least an agent $k\in\VV$ with $\bar{\mathcal{A}}^k\neq\{\}$ and an agent $j\in\VV$ with $\underline{\mathcal{A}}^j\neq\{\}$ ($k=j$ is possible). Therefore, for $l\in\bar{\mathcal{A}}^k$ it follows from~\eqref{eq::KKT_a3_s} that $\nu^\star=\frac{1}{\mathsf{w}^k_l}(-\nabla f^k_l(\mathsf{x}^{k\star }_l)-\bar{\mu}^{k\star}_l)$, and for $\bar{l}\in\underline{\mathcal{A}}^j$ it follows from~\eqref{eq::KKT2_a_s} that $\nu^\star=\frac{1}{\mathsf{w}^j_{\bar{l}}}(-\nabla f^j_{\bar{l}}(\mathsf{x}^{j\star }_{\bar{l}})+\bar{\mu}^{j\star}_{\bar{l}})$. Consequently, because $\bar{\mu}^{k\star}_l\geq0$ and $\bar{\mu}^{j\star}_{\bar{l}}\geq0$, we conclude that $-\frac{1}{\mathsf{w}^j_{\bar{l}}}\nabla f^j_{\bar{l}}(\mathsf{x}^{j\star }_{\bar{l}})\leq \nu^\star\leq -\frac{1}{\mathsf{w}^k_l}\nabla f^k_l(\mathsf{x}^{k\star }_l)$, which leads to
$|\nu^\star|\leq \max\{|\frac{\nabla f^j_{\bar{l}}(\mathsf{x}^{j\star }_{\bar{l}})}{{\mathsf{w}}^j_{\bar{l}}}|,|\frac{\nabla f^k_l(\mathsf{x}^{k\star }_l)}{{\mathsf{w}}^k_l}|\}\leq\frac{\max\{\|\nabla f^i(\vectsf{x}^{i\star})\|_\infty\}_{i=1}^N}{\underline{\mathsf{w}}}$. Therefore, we conclude that for any given $(\xs,\nu^\star,\{\bunderline{{\mu}}_l^{i\star}\}_{l\in\underline{\mathcal{B}}^i},\{\bar{{\mu}}_l^{i\star}\}_{l\in\bar{\mathcal{B}}^i})$, we have
$
|\nu^\star|\leq\frac{\max\{\|\nabla f^i(\vectsf{x}^{i\star})\|_\infty\}_{i=1}^N}{\underline{\mathsf{w}}}\leq  \frac{\max\big\{\underset{\vect{x}^i\in X^i_{\text{ineq}}}{\max}\,{\|\nabla f^i(\vect{x}^i)\|_\infty}\big\}_{i=1}^N}{{\bunderline{\mathsf{w}}}}
$. Consequently, it follows from~\eqref{eq::KKT_a3_s}  that  $\bar{\mu}^{i\star}_l\leq|\nabla f^i_l(\mathsf{x}^{i\star }_l)|\!
+|\mathsf{w}^i_l\,\nu^\star|\leq \|\nabla f^i_l(\mathsf{x}^{i\star }_l)\|_\infty+\bar{\mathsf{w}}|\nu^\star|$, and from~\eqref{eq::KKT2_a_s} that $\bunderline{\mu}^{i\star}_l\leq\|\nabla f^i_l(\mathsf{x}^{i\star }_l)\|_\infty\!
+|\mathsf{w}^i_l\,\nu^\star|\leq \|\nabla f^i_l(\mathsf{x}^{i\star }_l)\|_\infty+\bar{\mathsf{w}}|\nu^\star|$. Therefore, given~\eqref{eq::activeset_bound}, we have the guarantees that~\eqref{eq::mu_bound_p1} holds. 
\end{pf}
To compute the upper-bound in~\eqref{eq::mu_bound_p1} in a distributed manner, agents can run a set of max-consensus algorithms.  

To demonstrate the tightness of the bound in~\eqref{eq::inequl-multiplier-bound}, consider the following numerical example
  \begin{align*}
&\vectsf{x}^\star=\arg\min_{\vect{x}\in
    \real^{10}} \,\,\sum\nolimits_{i=1}^{10} f^i({x}^i),~~\text{subject~to~} \\
& ~w_1x^1+w_2x^2+\cdots+w_{10}x^{10}=b,\quad 0\leq x^i\leq1 \,\, i\in\mathbb{Z}_1^{10},
\end{align*}  
in which the local cost functions are assumed quadratic as $f^i({x}^i)=\alpha_i{x}^{i2}+\beta_i{x}^i+\gamma_i$ where the parameters chosen randomly according to $\alpha_i \in(0,1]$, $\beta_i \in(0,3]$, $\gamma_i \in(0,4]$, $b \in(0,4]$. The affine constraint weights are also chosen randomly according to $w_i\in(0,2]$ are randomly chosen.  For this problem finding the exact value of the Lagrange multipliers is possible by solving the KKT equations. To do this calculation, we  use fmincon function of MATLAB to obtain the optimum solution. Then, we compute the corresponding Lagrange multipliers by solving the KKT conditions. Table.~\ref{table.1} shows the values of $\mu_{max}$, the maximum of the Lagrange multipliers, and the values of $\mu_{bound}$ in~\eqref{eq::mu_bound_p1} 
for five different runs of the algorithm. As we can see, for this problem the values for $\mu_{bound}$ at most are only one order of magnitude larger than $\mu_{\max}$. 
{\small
\begin{table}[h]
   \caption{ {\small The  values of  actual $\mu_{bound}$ and the bound in~(17)}
    }
    \label{table.1}
    \centering
   \begin{tabular}{ |c|c|c|c|c|c|c|}
 \hline
 case: & 1 & 2 & 3 & 4 & 5 \\
 \hline
 $\mu_{max}$ 
  & 2.33  
  & 2.68  
  & 1.95   
  & 2.38
  & 1.95 \\
 \hline
  $\mu_{bound}$ in~(17) 
  & 13.34 
  & 17.91   
  & 11.6   
  & 52.1   
  & 18.48\\
   \hline
 \end{tabular}
\end{table}}

 Evaluating the MFCQ condition generally is challenging for other classes of optimization problems. 
  A common sufficient condition for the MFCQ is the linear independence constraint qualification (LICQ), which also guarantees the uniqueness of the Lagrange multipliers for any solution of the optimization 
problem~\eqref{eq::prob_def}~\cite{GW:13} (see~\cite{JW-NE:11} and \cite{PS-JC:18} for examples of the optimization solvers that are developed under the assumption that the LICQ holds). 
For a constrained optimization problem we say that the  LICQ holds for the optimal solution $\vectsf{x}^\star\in\real^m$ if the gradient of the equality constraints and the active inequality constraints at $\vectsf{x}^\star$ are linearly independent.  
The following result finds a ${\mu}_{\text{bound}}$ for problem~\eqref{eq::prob_def} when LICQ condition holds at the minimizers.


\begin{thm}\longthmtitle{Bounds on the Lagrange multipliers corresponding to inequality constraints when the LICQ holds at the minimizers}\label{eq::penalty-bound}
{\rm Consider  problem~\eqref{eq::prob_def} under Assumption~\ref{eq::assump_main-1}. Assume also that the LICQ holds at the minimizers of~\eqref{eq::prob_def}. Let $(\xs,\nus,\{\bunderline{{\mu}}_l^{i\star}\}_{l\in\underline{\mathcal{B}}^i},\{\bar{{\mu}}_l^{i\star}\}_{l\in\bar{\mathcal{B}}^i})$ be an arbitrary solution of the KKT conditions~\eqref{eq::KKT_1} for this problem. Then, the bound ${\mu}_{\text{bound}}$ in~\eqref{eq::mu_bound} satisfies
\begin{align}
&{\mu}_{\text{bound}} \!\leq\!\left(\!1\!+\!\frac{\bar{\mathsf{w}}}{\omega}\right)\!\max\big\{\underset{\vect{x}^i\in X^i_{\text{ineq}}}{\max}\,{\|\nabla f^i(\vect{x}^i)\|_\infty}\big\}_{i=1}^N.\label{eq::inequl-multiplier-bound}
\end{align}
where 
$\bar{\mathsf{w}}\!=\!\|\vectsf{W}\|_{\max}=\max\{\|\vectsf{w}^i\|_{\max}\}_{i=1}^N$,
and $\omega\!=\!\min\{\sigma_{\min}(\vectsf{W}_c)\,\big|\, \vectsf{W}_c\!\in\!\mathpzc{Q}(\vectsf{W}^\top)\, \}$. Here,  $\mathpzc{Q}(\vectsf{W}^\top)$ is the set of all the invertible $p\times p$ sub-matrices of $\vectsf{W}^\top\in\real^{m\times p}$ (recall~\eqref{eq::W_def}).}
\end{thm}
\begin{pf}
For any $(\xs,\nus,\{\bunderline{{\mu}}_l^{i\star}\}_{l\in\underline{\mathcal{B}}^i},\{\bar{{\mu}}_l^{i\star}\}_{l\in\bar{\mathcal{B}}^i})$, we note that the KKT conditions~\eqref{eq::KKT_1} can be written as 
\begin{subequations}\label{eq::KKT2}
\begin{align}
&\nabla f^i_l(\mathsf{x}^{i\star }_l)\!+\!([\vectsf{w}^i]^l)^\top\nus=0,\,\,\, l\in\mathbb{Z}_1^{n^i}\backslash\{\bar{\mathcal{A}}^i\cup{\bunderline{\mathcal{A}}^i}\},\label{eq::KKT_a2}\\
&\nabla f^i_l(\mathsf{x}^{i\star }_l)\!
+([\vectsf{w}^i]^l)^\top\nus+\bar{\mu}^{i\star}_l=0,\,\,\quad\quad l\in\bar{\mathcal{A}}^i,\label{eq::KKT_a3}\\
&\nabla f^i_l(\mathsf{x}^{i\star }_l)\!+([\vectsf{w}^i]^l)^\top\nus-\bunderline{\mu}^{i\star}_l=0,\,\,~~~\quad l\in\bunderline{\mathcal{A}}^i,\label{eq::KKT_a4}\\
&\sum\nolimits_{i=1}^{N}\sum\nolimits_{l=1}^{n^i}\,[\vectsf{w}^i]^l_j\,\mathsf{x}_l^{i\star }=\bs_j,\quad\quad\quad\quad
    j\in\mathbb{Z}_{1}^{p},\label{eq::KKT2_a} \\
&\mathsf{x}_l^{i\star } =\bunderline{\mathsf{x}}_l^i,~\,\qquad\qquad\qquad \qquad\qquad\qquad\,\,\, l\in\bunderline{\mathcal{A}}^i,\label{eq::prob_def-aug222_c}\\
&\mathsf{x}_l^{i\star }=\bar{\mathsf{x}}_l^i,\qquad\qquad\qquad\qquad\qquad\qquad\,\,\,\, ~l\in\bar{\mathcal{A}}^i,\label{eq::prob_def-aug222_d}
\end{align}
 \end{subequations}
 $i\in\VV$. Under the LICQ assumption, the gradients of the equality constraints (set of $p$ vectors in $\real^{m}$) and the active inequality constraints (set of $\sum_{i=1}^N|\underline{\mathcal{A}}^i\cup\bar{\mathcal{A}}^i|$ vectors in $\real^{m}$) at the minimizer should be linearly independent. This necessitates that $\sum_{i=1}^N|\underline{\mathcal{A}}^i\cup\bar{\mathcal{A}}^i|\leq m-p$. 
 As a result, we can conclude that $q=\sum_{i=1}^N|\mathbb{Z}_1^{n^i}\backslash(\bar{\mathcal{A}}^i\cup{\bunderline{\mathcal{A}}^i})|\geq p$. Thus, the number of KKT equations of the form~\eqref{eq::KKT_a2} is $q\geq p$. As a result, we can write all these $q$ equations as
 $$\vectsf{W}_e^\top\nus=-[\{\{\nabla f^i_l(\mathsf{x}^{i\star }_l)\}_{l=1}^{n^i}\}_{i=1}^N]$$
 where $\vectsf{W}_e\in\real^{p\times q}$ is a sub-matrix of $\vectsf{W}\in\real^{p\times m}$.
 Recall that under the LICQ assumption $(\nus\in\real^p,\{\bunderline{{\mu}}_l^{i\star}\}_{l\in\underline{\mathcal{B}}^i},\{\bar{{\mu}}_l^{i\star}\}_{l\in\bar{\mathcal{B}}^i})$ corresponding to every $\xs$ is unique. Thus,  $\rank(\vectsf{W}_e^\top)=p$ and there always exist a sub-matrix $\vectsf{W}_{se}\in\real^{p\times p}$ of $\vectsf{W}_e^\top\in\real^{q\times p}$ such that 
\begin{align}\label{eq::nubound2}
  \nus=-\vectsf{W}_{se}^{-1} \vect{J},
\end{align} 
 where $\vect{J}$ is the  components of $[\{\{\nabla f^i_l(\mathsf{x}^{i\star }_l)\}_{l=1}^{n_i}\}_{i=1}^N]$ associated with the rows of $\vectsf{W}_{se}$. Therefore, we can write
\begin{align}\label{eq::nus_bound}
\|\nus\|_{\infty}\!
\leq&\frac{1}{\sigma_{min}(\vectsf{W}_{se})}\|\vect{J}\|_\infty\nonumber\\
\leq& \frac{1}{\omega}\max\big\{\underset{\vect{x}^i\in X^i_{\text{ineq}}}{\max}\,{\|\nabla f^i(\vect{x}^i)\|_\infty}\big\}_{i=1}^N,
\end{align}
where $\omega$ is defined in the statement. Here, we used $|\nabla f_l^i(\mathsf{x}^{i\star})|\!\leq\! \max\big\{\!\!\underset{\vect{x}^i\in X^i_{\text{ineq}}}{\max}{\|\nabla f^i(\vect{x}^i)\|_\infty}\big\}_{i=1}^N$, $l\!\in\!\mathbb{Z}_{1}^{n^i}$, $i\!\in\!\VV$. On the other hand, given~\eqref{eq::KKT_a3} and~\eqref{eq::KKT_a4} we can write 
\begin{align*}
&\max\big\{\max\{\bunderline{\mu}_l^{i\star }\}_{l\in\underline{\mathcal{A}}^i},\max\{\bar{\mu}_l^{i\star }\}_{l\in\bar{\mathcal{A}}^i}\big\}_{i=1}^{N}\leq\\
&\quad \qquad 
 \max\big\{\underset{\vect{x}^i\in X^i_{\text{ineq}}}{\max}\,{\|\nabla f^i(\vect{x}^i)\|_\infty}\big\}_{i=1}^N\!+\bar{\mathsf{w}}\,\|\nus\|_\infty,
\end{align*}
where $\bar{\mathsf{w}}$ is defined in the statement. Therefore, given~\eqref{eq::nus_bound} we have the guarantees that~\eqref{eq::inequl-multiplier-bound} holds. 
\end{pf}

\section{Numerical examples}\label{sec::num}
\vspace{-0.1in}
In what follows, we demonstrate the performance of algorithm~\eqref{eq::Alg} via two numerical examples.

\begin{figure}[t!]
  \unitlength=0.5in
  \centering 
 \includegraphics[scale=0.6]{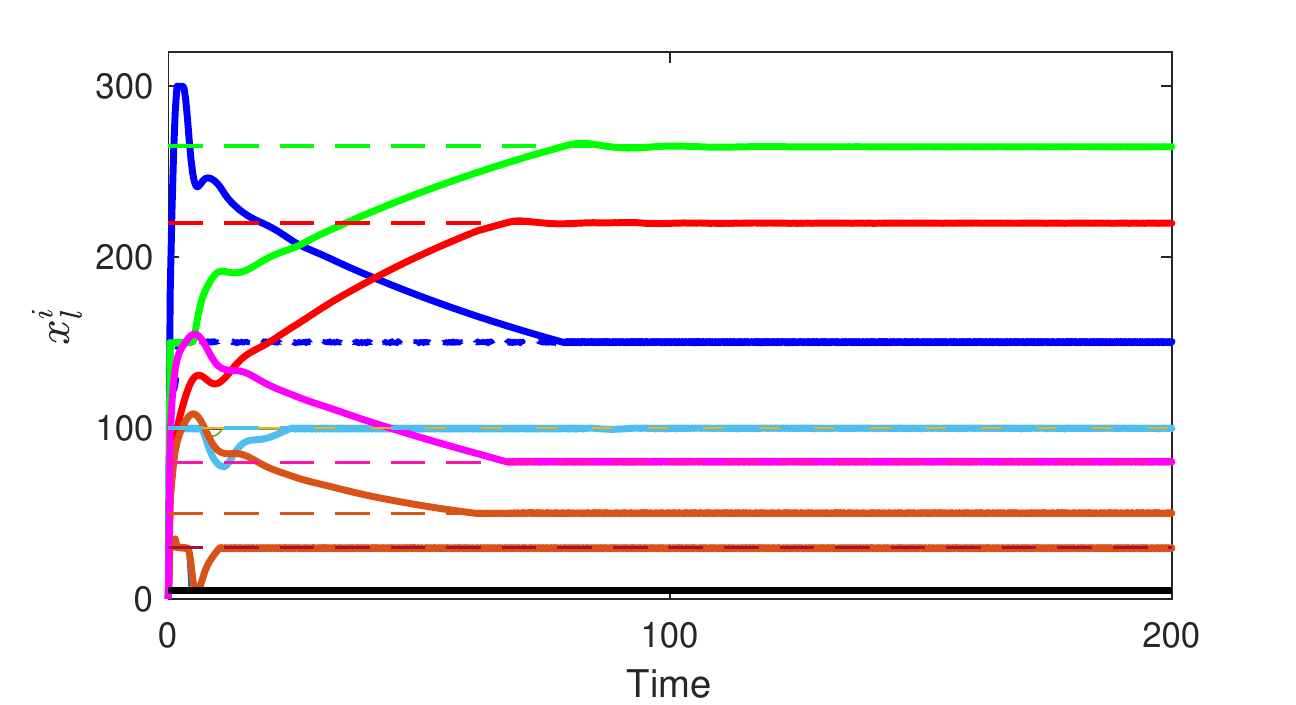}
   \caption{\scriptsize Execution of
   algorithm~\eqref{eq::Alg} over the network depicted in Fig.~\ref{fig::network}. The colored solid curved plots depicts the time history of decision variable of each agent. Horizontal  dashed lines depict the centralized solution obtained using MATLAB's constraint optimization solver `fmincon'.
    }\label{fig::Ex1sim_a}
\end{figure}

\begin{figure}[t!]
  \unitlength=0.5in
  \psfrag*{x}[][cc][1.1]{\renewcommand{\arraystretch}{0.5}\begin{tabular}{c}$\ln(|x^i-\mathsf{x}^\star|)$\\~\end{tabular}}
  \psfrag*{A}[][cc][1.1]{\renewcommand{\arraystretch}{0.5}\begin{tabular}{c}Agents\\~\end{tabular}}
  \psfrag*{t}[][cc][1.1]{\renewcommand{\arraystretch}{2}\begin{tabular}{c}$t$\end{tabular}}
  \centering {
    \includegraphics[height=1.2in]{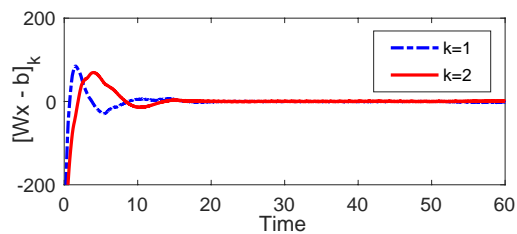}
  }
  \caption{{\scriptsize Constraint violation error while solving the optimization problem described in Figure~\ref{fig::network} using 
    algorithm~\eqref{eq::Alg}. }
    }
    \label{fig::Ex1b}
\end{figure}

As a first demonstrative example, we consider the in-network resource allocation problem described in Fig.~\ref{fig::network}. We choose the parameters of the costs and the limits of generation of the generators randomly from the table below, which lists the parameters of the generators of the IEEE 118 bus test model~\cite{IEEE-118},  located at buses $(4, 10, 18,26,54,69)$.
\begin{center}
{\tiny
\begin{tabular}{c*{5}{c}}
\hline\vspace{-0.1in}\\
\!IEEE\!\!             & $\alpha$ & $\beta$ & $\gamma$&  $\underline{\mathsf{x}}$  & $\bar{\mathsf{x}} $\\
bus number& [mu/MW$^2$] & [mu/MW] & [mu]&  [MW]  & [MW]\\
\hline
$4$ & 0.0696629 & 26.24382 & 031.67& 5 & 30   \\
$10$  &  0.010875 &12.8875&  6.78 & 150&  300  \\
$18$   &    0.0128 &17.82& 10.15 & 25& 100   \\
$26$  &  0.003 &10.76 &32.96 &100&350 \\
$54$&0.0024014 &12.32989 &28& 50&250\\
$69$& 0.010875 &12.8875& 6.78 &80&300\\\hline
\end{tabular}}
\end{center}
Figure~\ref{fig::Ex1sim_a} shows the time history of $x^i_l$'s generated by implementing the distributed optimization algorithm~\eqref{eq::Alg} (using $f^i_{\text{p}}(\vect{x}^i)$ as defined in~\eqref{eq::smooth_penalty} in place of $f^i(\vect{x}^i)$ in~\eqref{eq::Alg-x}) in comparison to the solution obtained using MATLAB's constraint optimization solver `fmincon'. As expected the decision variable $\vect{x}^i$ of each agent $i\until{6}$ converges closely to its corresponding minimizer, using $\eps=0.001$. Figure~\ref{fig::Ex1b} depicts the equality constraint violation time history, which as shown  vanishes over the time. 
For this problem to generate the dual dynamics, the agents $\{1,\cdots,6\}$, maintain and communicate variables of order $\{1,1,2,2,1,1\}$, respectively when we implement algorithm~\eqref{eq::Alg}. Whereas, if we implement algorithms of~\cite{LX-SB:06,YZ-MMZ:18,SAA-KY-AHS:18,PY-YH-FL:16,SSK:17,DD-MRJ:18,AC-JC:16-auto,AC-JC:15-tcns}, the corresponding variables to generate the dual dynamics is of order $\{4,2,6,6,4,2\}$.

\begin{figure}[t!]
  \unitlength=0.5in
  \centering 
    \includegraphics[scale=0.48]{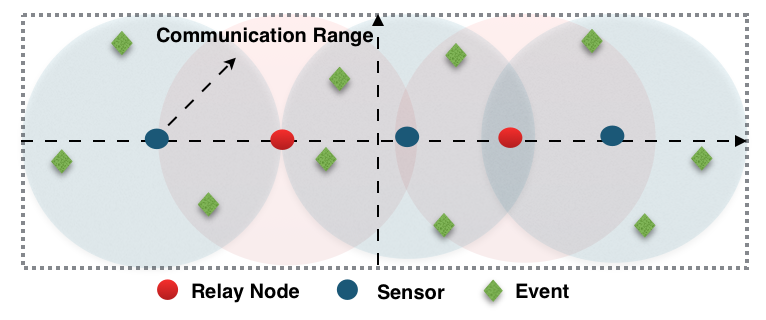}
   \caption{{\scriptsize Schematic representation of the events, sensors and relay nodes in the second example. }
   }\label{fig::Ex2}
\end{figure}

For second example, we consider a simple distributed self-localizing deployment problem concerned with optimal deployment of $3$ sensors labeled $\text{S}^i$, $i\in\{1,3,5\}$ on a line to monitor a set of events that are horizontally located at $\vect{P}\!=\![\{p_i\}_{i=1}^{10}]=[12,11,9,3,2,-1,-2,-8,-11,\\-13]$ for $t\in[0,100)$ , and $\vect{P}\!=\![\{p_i\}_{i=1}^{10}]\!=\![24,22,17,15,\\13,8,7,3,-2,-4]$ for $t\in[100,200)$, see Fig.~\ref{fig::Ex2}. Agent $1$ is monitoring $\{p_i\}_{i=1}^3$, agent $3$ is monitoring $\{p_i\}_{i=4}^7$, and agent $5$ is monitoring $\{p_i\}_{i=8}^{10}$.  Sensors should find their positions cooperatively to keep their position in the communication range of each other as well as stay close to the targets to improve the detection accuracy. Due to limited communication range, two relay nodes $\text{R}^i$, $i\in\{2,4\}$, as shown in Fig.~\ref{fig::Ex2} are used to guarantee the connectivity of the sensors during the operation.  The problem is formulated by
\begin{align}\label{eq::example2}
 \xs=&\arg\min_{\vect{x}\in
    \reals^{5}} \,\,\sum\nolimits_{i=1}^5 f^i(x^i),~~\text{subject~to~} \\
&~~ ~x^j-x^{j+1}\leq5,\quad\quad j\in\{1,\cdots,4\},\nonumber
\end{align}
where $f^i(x^i)=\sum_{j\in E^i}\|x^i-p_j\|^2$ for $i\in\{1,3,5\}$ with $E^1=\{1,\cdots,3\}$, $E^3=\{4,\cdots,7\}$ and $E^5=\{8,\cdots,10\}$ and $f^i(x^i)=0$ for $i\in\{2,4\}$. Here, $x^i$ with $i\in\{1,3,5\}$ (resp. $i\in\{2,4\}$) is the horizontal position of sensor $\text{S}^i$ (resp. relay node $\text{R}^i$). To transform problem~\eqref{eq::example2} to the standard form described in~\eqref{eq::prob_def} we introduce slack variables $x^i_2\in\real$ with $i\in\{1,\cdots,4\}$, to rewrite~\eqref{eq::example2} as
\begin{align}\label{eq::example2_eq}
&\xs=\arg\min_{\vect{x}\in
    \reals^{9}} \,\,\sum\nolimits_{i=1}^5 f^i(\vect{x}^i),~~\text{subject~to~} \\
& ~x^j_1-x^{j+1}_1+x^{j}_2=5,~~ x^{j}_2\geq0,~~ j\in\{1,\cdots,4\}, \nonumber 
\end{align}
where $\vect{x}^i\in\real^2$ for $i\in\{1,2,3,4\}$, $\vect{x}^5\in\real$, and $f^i(\vect{x}^i)=f^i(x^i_1)$ for any $i\in\{1,\cdots,5\}$, i.e., $f^i(x^i_2)=0$.
We can run  algorithm~\eqref{eq::Alg} by choosing the cyber layer equivalent to the physical connected topology between all the agent, i.e., $\GG_k=\GG$ for $k\in\{1,2,3,4\}$, where $\GG$ is the line graph connecting all $5$ agents.  However, as stated earlier this configuration leads to extra computational and communication efforts. Here, instead, we  form $4$ cyber-layers $\GG_k$, $k\in\{1,2,3,4\}$, where $\VV_1=\{1,2\}$, $\VV_2=\{2,3\}$, $\VV_3=\{3,4\}$ and $\VV_4=\{4,5\}$. We note  that our  proposed approach to form the cyber-layers in correspondence to the equality constraints leads to an efficient communication topology here. More specifically, to generate the dual dynamics, the agents $\{1,\cdots,5\}$, maintain and communicate variables of order $\{1,2,2,2,1\}$, respectively. Whereas, if we implement algorithms of~\cite{YZ-MMZ:18,SSK:17}, 
the corresponding variables to generate the dual dynamics is of order $\{8,8,8,8,4\}$.

\begin{figure}[t!]
  \unitlength=0.5in
  \centering 
        \includegraphics[trim=15pt 0 0 0,clip,scale=0.6]{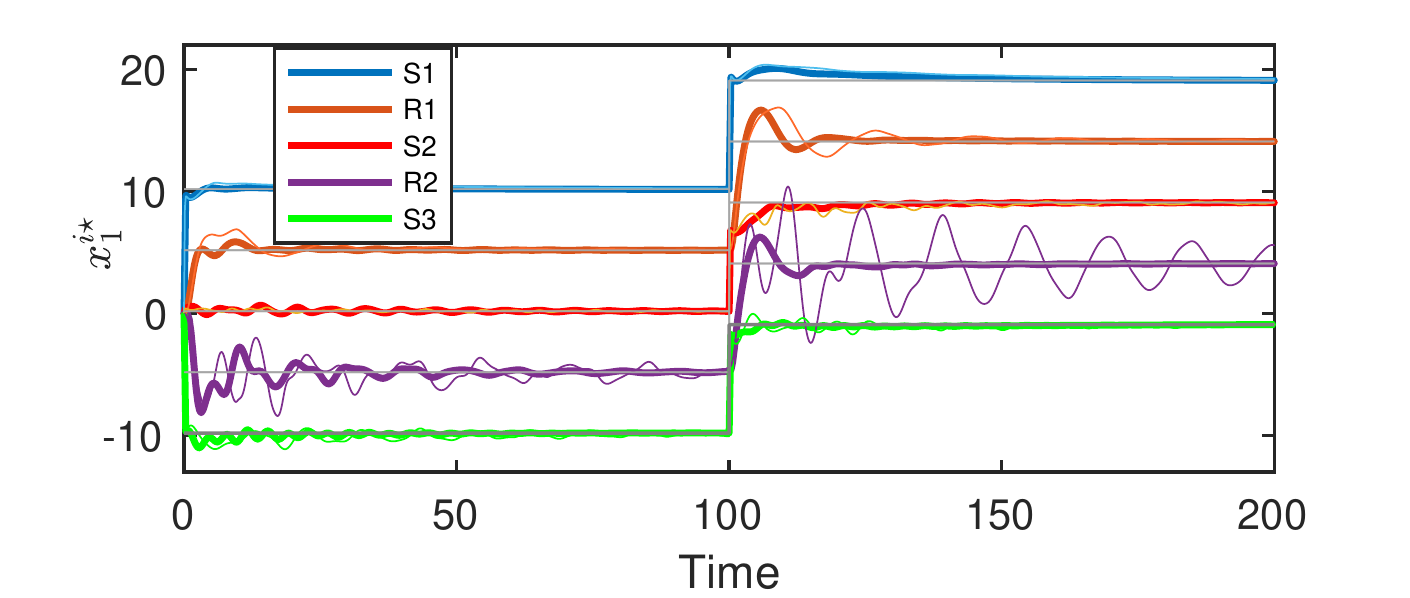}
   \caption{\scriptsize Trajectories of $\{x_1^i\}_{i=1}^5$ generated by implementing distributed algorithm~\eqref{eq::Alg}: The grey lines show the optimum positions of agents on the line obtained by using the Matlab's fmincon.  
   The thick curved lines show the trajectories when algorithm~\eqref{eq::Alg} is implemented over cluster-based cyber-layers. the thin lines show the trajectories when algorithm~\eqref{eq::Alg} is implemented with $\mathcal{G}_k=\mathcal{G}$, $k=\{1,2,3,4\}$. 
    }\label{fig::ex2_plot}
\end{figure}

  \vspace{-0.05in}
Figure~\ref{fig::ex2_plot} shows the trajectory of the distributed optimization algorithm~\eqref{eq::Alg} (using $f^i_{\text{p}}(\vect{x}_1^i)$ as defined in~\eqref{eq::smooth_penalty} in place of $f^i(\vect{x}^i)$ in~\eqref{eq::Alg-x}) for problem~\eqref{eq::example2_eq}. As shown the location of the sensors remain in their communication range and converge to optimum values during execution of the algorithm (the optimal solution is shown by the grey lines, and is obtained by  MATLAB's constraint optimization solver `fmincon'). 
Our choice of smooth penalty function~\eqref{eq::smooth_penalty} is obtained by $\gamma=200$ and $\epsilon=0.01$ which satisfies the condition of Proposition~\ref{prop::penalty}.
What is interesting to note in Fig.~\ref{fig::ex2_plot} is how the convergence of the algorithm is slowed down when we use $\GG_k=\GG$ for $k\in\mathbb{Z}_1^4$. This is expected, as in this case the coordination to generate the dual variables has to happen over a larger graph. 

Table~\ref{table::eps_feas_bound} gives the global cost value and the inequality constraint evaluation at $\vectsf{x}_p^\star$ obtained by using our distributed algorithm with $\eps$-exact penalty function method for three simulation scenarios. The first and the second scenarios are respectively when we use $\eps=0.01$ and $\eps=0.001$. As we can see when $\eps=0.01$ only one of the inequalities is violated slightly (by $2.1\text{e}^{-4}$). When a smaller $\eps=0.001$ is used this violation also is removed. Table~\ref{table::eps_feas_bound} also shows that if we use the 'adjusted boxed inequalities' that we introduced in Remark~\ref{rem::adjusted}, the inequality constraints are all respected with only a negligible increase in the cost value. 

\begin{table*}[h]
    \caption{\small The global cost value and the inequality constraint evaluation at $\vectsf{x}_p^\star$ obtained by using $\eps$-exact penalty function method }\label{table::eps_feas_bound}
    \centering
   \begin{tabular}{ |c|c|c|c|c|c|c| } 
 \hline
  & $x^1-x^2-5$& $x^2-x^3-5$ & $x^3-x^4-5$ & $x^4-x^5-5$ &  $f(\vect{x}_p^\star)$\\ 
 \hline
 $\eps=0.01$ 
  & -5.8e-3  
  & -2.06e-2
  & -2.63e-2   
  & \color{purple}{2.1e-4}
  & 680.4 
 \\ 
 \hline
  $\eps=0.001$  
  &  -5.92-3
  &  -3.46e-2  
  &  -3.75e-2 
  &   -3.5e-2 
  & 680.4
 \\ 
 \hline
  $\eps=0.01$ and adjusted bounds  
  & -1.25e-2
  & -1.3e-2  
  &  -3.92e-2  
  &   -8.2e-3 
  & 680.23
\\ 
   \hline
 \end{tabular}
\end{table*}


\section{Conclusions}\label{sec::conclude} 
\vspace{-0.13in}
We proposed a novel cluster-based distributed augmented Lagrangian algorithm for a class of constrained convex optimization problem. 
In the design of our distributed algorithm, we paid special attention to the efficient communication and computation resource management and required only the agents that are coupled through an equality constraint to form a communication topology to address that coupling in a distributed manner.
 We showed that if the communication topology corresponding to each equality constraint is a connected graph, the proposed algorithm converges asymptotically when the local cost functions are convex, and exponentially when the local cost functions are strongly convex and have Lipschitz gradients.  We invoked the $\eps$-exact penalty function method to address the inequality constraints and obtained an explicit lower bound on the penalty function weight to guarantee convergence to $\eps$-neighborhood of the global minimum value of the cost. Simulations demonstrated the performance of our proposed algorithm.
  As future work, we will study the event-triggered communication implementation of our~algorithm.

\vspace{-0.13in}
\bibliographystyle{ieeetr}%
\bibliography{bib/alias,bib/Reference} 

\begin{thebibliography}{10}

\bibitem{SSK:17acc}
S.~S. Kia, ``An augmented lagrangian distributed algorithm for an in-network
  optimal resource allocation problem,'' in {\em {A}merican {C}ontrol
  {C}onference}, (WA, USA), 2017.

\bibitem{AJW-FW-GBS:13}
A.~J. Wood, F.~Wollenberg, and G.~B. Sheble, {\em Power Generation, Operation
  and Control}.
\newblock New York: John Wiley, 3rd~ed., 2013.

\bibitem{AC-JC:16}
A.~Cherukuri and J.~Cortés, ``Initialization-free distributed coordination for
  economic dispatch under varying loads and generator commitment,'' {\em
  Automatica}, vol.~74, pp.~183--193, 2016.

\bibitem{LX-M-SPB:04}
L.~Xiao, M.~Johansson, and S.~P. Boyd, ``Simultaneous routing and resource
  allocation via dual decomposition,'' {\em IEEE Transactions on
  Communications}, vol.~52, no.~7, pp.~1136--1144, 2004.

\bibitem{RM-SL:06}
R.~Madan and S.~Lall, ``Distributed algorithms for maximum lifetime routing in
  wireless sensor networks,'' {\em IEEE Transactions on Wireless
  Communications}, vol.~5, no.~8, pp.~2185--2193, 2006.

\bibitem{JC-VKNL:12}
J.~Chen and V.~K.~N. Lau, ``Convergence analysis of saddle point problems in
  time varying wireless systems -- control theoretical approach,'' {\em IEEE
  Transactions on Signal Processing}, vol.~60, no.~1, pp.~443--452, 2012.

\bibitem{AF-FP:14}
A.~Ferragut and F.~Paganini, ``Network resource allocation for users with
  multiple connections: fairness and stability,'' {\em IEEE/ACM Transactions on
  Networking}, vol.~22, no.~2, pp.~349--362, 2014.

\bibitem{SAA-KY-AHS:18}
S.~A. Alghunaim, K.~Yuan, and A.~H. Sayed, ``Dual coupled diffusion for
  distributed optimization with affine constraints,'' in {\em {IEEE} Conf. on
  Decision and Control}, (FL, USA), 2018.

\bibitem{RR-GC-DG:17}
R.~Rostami, G.~Costantini, and D.~Görges, ``{A}{D}{M}{M}-based distributed
  model predictive control: Primal and dual approaches,'' in {\em {IEEE} Conf.
  on Decision and Control}, (Melbourne, Australia), 2017.

\bibitem{SB-NP-EC-BP-JE:10}
S.~Boyd, N.~Parikh, E.~Chu, B.~Peleato, and J.~Eckstein, ``Distributed
  optimization and statistical learning via the alternating direction method of
  multipliers,'' {\em Foundations and Trends in Machine Learning}, vol.~3,
  pp.~1--122, 2010.

\bibitem{JD-AA-MW:12}
J.~Duchi, A.~Agarwal, and M.~Wainwright, ``Dual averaging for distributed
  optimization: {C}onvergence analysis and network scaling,'' {\em IEEE
  Transactions on Automatic Control}, vol.~57, no.~3, pp.~592--606, 2012.

\bibitem{JW-NE:11}
J.~Wang and N.~Elia, ``A control perspective for centralized and distributed
  convex optimization,'' in {\em {IEEE} Conf. on Decision and Control}, (FL,
  USA), 2011.

\bibitem{SSK-JC-SM:15-auto}
S.~S. Kia, J.~Cort{\'e}s, and S.~Mart{\'\i}nez, ``Distributed convex
  optimization via continuous-time coordination algorithms with discrete-time
  communication,'' {\em Automatica}, vol.~55, pp.~254--264, 2014.

\bibitem{DV-FZ-AC-GP-LS:15}
D.~Varagnolo, F.~Zanella, A.~Cenedese, G.~Pillonetto, and L.~Schenato,
  ``Newton-raphson consensus for distributed convex optimization,'' {\em IEEE
  Transactions on Automatic Control}, vol.~61, no.~4, pp.~994 -- 1009, 2015.

\bibitem{ZZ-MC:12}
Z.~Zhang and M.~Chow, ``Convergence analysis of the incremental cost consensus
  algorithm under different communication network topologies in a smart grid,''
  {\em IEEE Transactions on Power Systems}, vol.~27, no.~4, pp.~1761--1768,
  2012.

\bibitem{SK-GH:12}
S.~Kar and G.~Hug, ``Distributed robust economic dispatch in power systems: A
  consensus + innovations approach,'' in {\em Power \& Energy Society General
  Meeting}, (San Diego, CA), pp.~1--8, July 2012.

\bibitem{ADDG-STC-CNH:12}
A.~D. Dominguez-Garcia, S.~T. Cady, and C.~N. Hadjicostis, ``Decentralized
  optimal dispatch of distributed energy resources,'' in {\em {IEEE} Conf. on
  Decision and Control}, (Hawaii, USA), pp.~3688--3693, Dec. 2012.

\bibitem{LX-SB:06}
L.~Xiao and S.~Boyd, ``Optimal scaling of a gradient method for distributed
  resource allocation,'' {\em Journal of optimization theory and applications},
  vol.~129, no.~3, pp.~469--488, 2006.

\bibitem{YZ-MMZ:18}
Y.~Zhang and M.~M. Zavlanos, ``A consensus-based distributed augmented
  lagrangian method,'' in {\em {IEEE} Conf. on Decision and Control}, (CA,
  USA), 2018.

\bibitem{PY-YH-FL:16}
P.~Yi, Y.~Hong, and F.~Liu, ``Initialization-free distributed algorithms for
  optimal resource allocation with feasibility constraints and its application
  to economic dispatch of power systems,'' {\em Automatica}, vol.~74,
  pp.~259--269, 2016.

\bibitem{SSK:17}
S.~S. Kia, ``Distributed optimal in-network resource allocation algorithm
  design via a control theoretic approach,'' {\em Systems and Control Letters},
  vol.~107, pp.~49--57, 2017.

\bibitem{DD-MRJ:18}
D.~Ding and M.~Jovanovic´, ``A primal-dual {L}aplacian gradient flow dynamics
  for distributed resource allocation problems,'' in {\em {A}merican {C}ontrol
  {C}onference}, (WI, USA), 2018.

\bibitem{AC-JC:16-auto}
A.~Cherukuri and J.~Cort{\'e}s, ``Initialization-free distributed coordination
  for economic dispatch under varying loads and generator commitment,'' {\em
  Automatica}, vol.~74, pp.~183--193, 2016.

\bibitem{AC-JC:15-tcns}
A.~Cherukuri and J.~Cort{\'e}s, ``Distributed generator coordination for
  initialization and anytime optimization in economic dispatch,'' {\em IEEE
  Transactions on Control of Network Systems}, vol.~2, no.~3, pp.~226--237,
  2015.

\bibitem{DPB-JNT:97}
D.~Bertsekas and J.~Tsitsiklis, {\em Parallel and Distributed Computation:
  Numerical Methods}.
\newblock 1997.

\bibitem{DJ-JMFM-JX:15}
D.~Jakovetic, J.~Moura, and J.~Xavier, ``Linear convergence rate of a class of
  distributed augmented lagrangian algorithms,'' {\em IEEE Transactions on
  Automatic Control}, vol.~60, no.~4, pp.~922--936, 2015.

\bibitem{MV-JC:18}
M.Vaquero and J.Cortes, ``Distributed augmentation-regularization for robust
  online convex optimization,'' {\em IFAC-PapersOnLine}, vol.~51, no.~23,
  pp.~230--235, 2018.

\bibitem{KJA-LH-HU:58}
K.~J. Arrow, L.~Hurwicz, and H.~Uzawa, {\em Studies in linear and nonlinear
  programming}.
\newblock 1958.

\bibitem{DD-BH-NKD-MRJ:18}
D.~Ding, B.~Hu, N.~Dhingra, and M.~Jovanovic´, ``An exponentially convergent
  primal-dual algorithm for nonsmooth composite minimization,'' in {\em {IEEE}
  Conf. on Decision and Control}, (FL, USA), 2018.

\bibitem{SSK:16-ifac}
S.~S. Kia, ``Distributed optimal resource allocation over networked systems and
  use of an epsilon-exact penalty function,'' in {\em IFAC Symposium on Large
  Scale Complex Systems}, (CA, USA), 2016.

\bibitem{WMH-VC:08}
W.~Haddad and V.~Chellaboina, {\em Nonlinear Dynamical Systems and Control}.
\newblock Princeton University Press, 2008.

\bibitem{DPB:75}
D.~P. Bertsekas, ``Nondifferentiable optimization via approximation,'' {\em
  {M}athematical {P}rograming {S}tudy}, vol.~3, pp.~1--25, 1975.

\bibitem{MCP-SAZ:94}
M.~\c{C}. Pinar and S.~A. Zenios, ``On smoothing exact penalty functions for
  convex constrained optimization,'' {\em IEEE Transactions on Communications},
  vol.~4, no.~3, pp.~1136--1144, 1994.

\bibitem{WW-JW-NL-SM:17}
W.~Wei, J.~Wang, N.~Li, and S.~Mei, ``Optimal power flow of radial networks and
  its variations: A sequential convex optimization approach,'' {\em IEEE
  Transactions on Smart Grid}, vol.~8, no.~6, pp.~2974--2987, 2017.

\bibitem{MZ-DF-ADD:19}
M.~Zholbaryssov, D.~Fooladivanda, and A.~D.Domínguez-García, ``Resilient
  distributed optimal generation dispatch for lossy ac microgrids,'' {\em
  Systems and Control Letters}, vol.~123, pp.~47--54, 2019.

\bibitem{OM:85}
O.~Mangasarian, ``Computable numerical bounds for lagrange multipliers of
  stationary points of non-convex differentiable non-linear programs,'' {\em
  Operations Research Letters}, vol.~4, no.~2, pp.~1757--1780, 1985.

\bibitem{SR-MM-CJ:11}
S.~Richter, M.~Morari, and C.~Jones, ``Towards computational complexity
  certification for constrained {M}{P}{C} based on {L}agrange relaxation and
  the fast gradient method,'' in {\em {IEEE} Conf. on Decision and Control},
  (Orlando, {F}lorida, USA), pp.~5223 -- 5229, 2011.

\bibitem{FB-JC-SM:09}
F.~Bullo, J.~Cort{\'e}s, and S.~Mart{\'\i}nez, {\em Distributed Control of
  Robotic Networks}.
\newblock Applied Mathematics Series, Princeton University Press, 2009.

\bibitem{SB-LV:04}
S.~Boyd and L.~Vandenberghe, {\em Convex Optimization}.
\newblock Cambridge University Press, 2004.

\bibitem{DPB:99}
D.~Bertsekas, {\em Nonlinear Programming}.
\newblock 1999.

\bibitem{OLM-SF:67}
O.~L. Mangasarian and S.~Fromovitz, ``The fritz john necessary optimality
  conditions in the presence of equality and inequality constraints,'' {\em
  Operations Research Letters}, vol.~17, pp.~37--47, 1967.

\bibitem{GW:13}
G.~Wachsmuth, ``On {LICQ} and the uniqueness of {L}agrange multipliers,'' {\em
  Operations Research Letters}, vol.~41, no.~1, pp.~78--80, 2013.

\bibitem{PS-JC:18}
P.~Srivastava and J.~Cortes, ``Distributed algorithm via continuously
  differentiable exact penalty method for network optimization,'' in {\em
  {IEEE} Conf. on Decision and Control}, (FL, USA), 2018.

\bibitem{IEEE-118}
2004.
\newblock \url{http://motor.ece.iit.edu/data/JEAS\_IEEE118.doc}.

\bibitem{HKK:02}
H.~K. Khalil, {\em Nonlinear Control}.
\newblock Prentice Hall, 2002.

\end{thebibliography}

\section*{Appendix A}
\vspace{-0.13in}
\renewcommand{\theequation}{A.\arabic{equation}}
\renewcommand{\thethm}{A.\arabic{thm}}
\renewcommand{\thelem}{A.\arabic{lem}}
\renewcommand{\thedefn}{A.\arabic{defn}}

\begin{pf} [Proof of Theorem~\ref{thm::main}]
   Let $(\{\vectsf{x}^{i\star}\}_{i=1}^N,\nus)$ satisfy the KKT equation~\eqref{eq::KKT_1} and  $\vect{y}^\star_k=[\{[\vectsf{w}^l]_k \vectsf{x}^{l\star}-\bar{\bs}^l_k\}_{l\in\VV_k}]$. For convenience in analysis, we apply the change of variables 
\begin{align}\label{eq::new_coord}
  \!  \!\vect{q}_k\!&=\!\begin{bmatrix}\rr_k^\top\!\\\rR_k^\top\!\end{bmatrix} \!(\vect{y}_k\!-\!\vect{y}^\star_k),\,\, \vect{p}_k\!=\!\vect{v}_k\!-\!\nu^\star_k\vect{1}_{N_k},\, \vect{\chi}^i=\vect{x}^i\!-\!\vectsf{x}^{i\star},
\end{align}
 to write the algorithm~\eqref{eq::Alg}, under the stated initialization conditions, in the equivalent form
 \begin{subequations}\label{eq::Alg_col}
  \begin{align}
      \dot{\hat{q}}_k  =&\, 0,\label{eq::Alg_q1}\\
      \overset{.}{\bar{\vect{q}}}_k=&\,\beta_k\, (\rR_k^\top\lL_k\rR_k)\,\rR_k^\top\vect{p}_k, \label{eq::Alg_col_q2}\\
    \dvect{p}_k  =&\, \vectsf{\psi}_k    \vect{\chi}_k-\beta_k\,\lL_k\vect{p}_k-\rR_k\,\bar{\vect{q}}_k-\rr_k\, \hat{q}_k,\label{eq::Alg_col_p}\\
      \dvect{\chi}^i=&\,-(\rho^i+1)\,(\nabla f^i(\vect{\chi}^i+\xsi)-\nabla f^i(\xsi))\,+\,\nonumber\\&\sum\nolimits_{k\in\mathcal{T}^i}\big(-\rho^i[\vectsf{w}^i]_{k}^{\top}[\vectsf{w}^i]_{k}\vect{\chi}^i
      \,-(\rho^i+1)[\vectsf{w}^{i}]^\top_kp_k^i
      \nonumber\\&+\rho^i{[\vectsf{w}}^{i}]^\top_k[\rR_k\,\bar{\vect{q}}_k]^i+\rho^i\,{[\vectsf{w}}^{i}]^\top_k\! \,\hat{q}_k\big),\label{eq::Alg_col_chi} 
  \end{align}
  \end{subequations}
  where we used $\vect{q}_k=(\hat{q}_k,\bar{\vect{q}}_k)$ with $\hat{q}_k\in\!\real,\, \bar{\vect{q}}_k\in\!\real^{(N_k-1)}$. Here, we also used $\rR_k\rR_k^\top\lL_k=\lL_k$, $\vectsf{\psi}_k=\text{Blkdiag}(\{[\vectsf{w}^i]_k\}_{i\in\VV_k})$ and
$\vect{\chi}_k=[\{\vect{\chi}^{i\top}\}_{i\in\VV_k}]^\top$.  Under the given initial condition, for any $t\in\realnonnegative$ we obtain 
   \begin{align}\label{eq::Th1_q1}\hat{q}_k(t)=\!\frac{1}{\sqrt{N_k}}\big(\sum\nolimits_{l\in\VV_k}\!\!y_k^l(t)\!-\!\!([\vectsf{W}]_k\xs\!-\!\bs_k)\big)\!=\!0.
   \end{align}
   To study the stability in the other variables, we let $\hat{q}_k(t)=0$ in \eqref{eq::Alg_col_p} and~\eqref{eq::Alg_col_chi}, and consider the radially unbounded candidate
  Lyapunov function 
  \begin{align}\label{eq::Lya_func}
   &V(\{\bar{\vect{q}}_k\}_{k=1}^p,\{\vect{p}_k\}_{k=1}^p,\{\vect{\chi}^i\}_{i=1}^N)=\frac{1}{2}\sum\nolimits_{i=1}^{N}\vect{\chi}^{i\top}\vect{\chi}^i\,+\nonumber\\&\quad\frac{1}{2} \sum\nolimits_{k=1}^{p}\big(\bar{\vect{q}}_k^{\top}(\vect{\Gamma}_k+\vectsf{I})(\beta_k\rR_k^\top\lL_k\rR_k)^{-1}\bar{\vect{q}}_k+\vect{p}^\top_k\vect{p}_k\nonumber\\&\qquad 
   +(\vect{p}_k+\rR_k\,\bar{\vect{q}}_k)^\top\vect{\Gamma}_k\,(\vect{p}_k+\rR_k\,\bar{\vect{q}}_k)\big),
   \end{align}
   where $\vect{\Gamma}_k\!=\!\text{Blkdiag}(\{\rho^i\}_{i\in\VV_k})$. Note that   $(\beta_k\rR_k^\top\lL_k\rR_k)^{-1}$ and  $\vect{\Gamma}_k+\vectsf{I}$ are positive definite diagonal matrices, thus  $\bar{\vect{q}}_k^{\top}(\vect{\Gamma}_k+\vectsf{I})(\beta_k\rR_k^\top\lL_k\rR_k)^{-1}\bar{\vect{q}}_k\!>\!\vect{0}$.  Taking the derivative of $V$ along the trajectories of~\eqref{eq::Alg_col_q2}-\eqref{eq::Alg_col_chi} gives
   \begin{align}\label{eq::dot_V_CG}
      &\!\dot{V}\!=\!-\!\!\sum\nolimits_{i=1}^{N}\!\big((\rho^i+1)\vect{\chi}^{i\top}(\nabla f(\vect{\chi}^i+\xsi)-\!\nabla f(\xsi))\\&\!
     -\!\sum_{k=1}^{p}\!\!\big(\beta_k\,\vect{p}_k^\top\lL_k\vect{p}_k\!+\!(\vectsf{\psi}_k\,\vect{\chi}_k\!-\!\rR_k\bar{\vect{q}}_k)^\top\!\vect{\Gamma}_k\!(\vectsf{\psi}_k\vect{\chi}_k\!-\!\rR_k\bar{\vect{q}}_k)\big)
      .\nonumber
   \end{align}
Convexity of the local cost functions ensures $\vect{\chi}^i(\vect{\nabla} f^i(\vect{\chi}^i+\xsi)-\vect{\nabla} f^i(\xsi))=((\vect{\chi}^i+\xsi)-\xsi)(\vect{\nabla} f^i(\vect{\chi}^i+\xsi)-\vect{\nabla} f^i(\xsi))\geq 0$, $i\in\VV$.
  The connectivity of the sub-graph $\GG_k$, $k\in\mathbb{Z}_1^p$ also ensures  $-\vect{p}_k^\top\lL_k\vect{p}_k\leq0$. Thus,  $\dot{V}\leq0$, and consequently  the trajectories of~\eqref{eq::Alg_col_q2}-\eqref{eq::Alg_col_chi} starting from any initial condition are bounded. 
  
  \vspace{-0.05in}
  Next, we invoke the invariant set stability results to prove that the trajectories of~\eqref{eq::Alg_col_q2}-\eqref{eq::Alg_col_chi} converge to a point in its set of equilibrium points. Let $\mathcal{S}
  =\{(\{\bar{\vect{q}}_k\}_{k=1}^p,\{\vect{p}_k\}_{k=1}^p,\{\vect{\chi}^i\}_{i=1}^N)\in\prod_{k=1}^p\!\real^{N_k-1}\times\prod_{k=1}^p\!\real^{N_k}\times\prod_{i=1}^N\reals^{n^i} |\,
  \dot{V}\equiv0\}$. Given~\eqref{eq::dot_V_CG}, we have $\mathcal{S}=
  \Big\{\!\{\bar{\vect{q}}_k\}_{k=1}^p,\!\{\vect{p}_k\}_{k=1}^p,\!\{\vect{\chi}^i\}_{i=1}^N\!\in\!\prod_{k=1}^p\real^{N_k-1}\!\times\!\prod_{k=1}^p\real^{N_k}\times\prod_{i=1}^N\reals^{n^i}\Big|
 ~\vect{p}_k=\vect{0},~\vectsf{\psi}_k\,\vect{\chi}_k=\rR_k\bar{\vect{q}}_k,\\
 \vect{\chi}^{i\top}(\nabla f^i(\vect{\chi}^i+\xsi)-\nabla f^i(\xsi))=0,~ i\in\VV,~k\in\mathbb{Z}_1^p\big\}$.
  Since $\vect{\chi}^{i\top}(\nabla f^i(\vect{\chi}^i+\xsi)-\nabla f^i(\xsi))=\sum_{j=1}^{n^i}\chi^i_j(\nabla f^i_j({\chi^i_j}+\mathsf{x}^{i\star }_j)-\nabla f^i_j(\mathsf{x}^{i\star }_j))$, due to convexity of the cost functions $f^i_j$, $j\in\mathbb{Z}_{1}^{n_i}$, $i\in\VV$, from $\vect{\chi}^{i\top}(\nabla f^i(\vect{\chi}^i+\xsi)-\nabla f^i(\xsi))=0$ we conclude that either ${\chi}_j^i=0$ or $\nabla f_j^i({\chi^i_j}+\mathsf{x}^{i\star }_j))-\nabla f_j^i(\mathsf{x}^{i\star }_j))=0$. Consequently, the points in $\mathcal{S}$ satisfy $\nabla f^i(\vect{\chi}^i+\xsi)-\nabla f^i(\xsi)=0$. As a result, given~\eqref{eq::Th1_q1}, a trajectory $t
  \mapsto (\{\bar{\vect{q}}_k(t)\}_{k=1}^p,\{\vect{p}_k(t)\}_{k=1}^p,\{\vect{\chi}^i(t)\}_{i=1}^N)$ of~\eqref{eq::Alg_col_q2}-\eqref{eq::Alg_col_chi} belonging to
  $\mathcal{S}$ for all $t\geq 0$, must satisfy $(\overset{.}{\bar{\vect{q}}}_k\equiv\vect{0},\dvect{p}_k\equiv \vect{0},\dvect{\chi}^i\equiv \vect{0})$.
 Therefore, the largest invariant set in $\mathcal{S}$ is the set of equilibrium points of~\eqref{eq::Alg_col_q2}-\eqref{eq::Alg_col_chi}. Then, invoking the La~Salle invariant theorem ~\cite[Theorem 3.4]{WMH-VC:08}, we conclude that the trajectories of~\eqref{eq::Alg_col_q2}-\eqref{eq::Alg_col_chi} converge asymptotically to the set of its equilibrium points. 
 
   \vspace{-0.05in}
 Next, we show that the convergence is indeed to a point in the equlibia set. For that, by virtue of semi-stability theorem \cite[Theorem 4.20]{WMH-VC:08}, we show that every equilibrium point of ~\eqref{eq::Alg_col_q2}-\eqref{eq::Alg_col_chi} is Lyapunov stable.  Let $(\{\bunderline{\vect{\bar{q}}}_k\}_{k=1}^p,\{\bunderline{\vect{p}}_k\},\{\bunderline{\vect{\chi}}^i)\}_{i=1}^N$ be an equilibrium point of ~\eqref{eq::Alg_col_q2}-\eqref{eq::Alg_col_chi} (recall that $\hat{q}_k(t)=0$ due to~\eqref{eq::Th1_q1}). Now, consider the change of variables  $\mathbf{\mathfrak{q}}_k=\bar{\vect{q}}_k-\bunderline{\vect{\bar{q}}}_k$ and $\mathbf{\mathfrak{p}}_k=\vect{p}_k-\bunderline{\vect{p}}_k$ for $k\in\mathbb{Z}_1^p$, and $\mathbf{\mathfrak{r}}^i=\vect{\chi}^i-\bunderline{\vect{\chi}}^i$ for $i\in\VV$, to write~\eqref{eq::Alg_col_q2}-\eqref{eq::Alg_col_chi} as
 \begin{subequations}\label{eq::Alg_col2}
  \begin{align}
    \dot{\mathbf{\mathfrak{q}}}_k=&\,\beta_k\, (\rR_k^\top\lL_k\rR_k)\,\rR_k^\top\mathbf{\mathfrak{p}}_k, \label{eq::Alg_col_q22}\\
    \dot{\mathbf{\mathfrak{p}}}_k  =&\, \vectsf{\psi}_k    \mathbf{\mathfrak{r}}_k-\beta_k\,\lL_k\mathbf{\mathfrak{p}}_k-\rR_k\,\mathbf{\mathfrak{q}}_k,\\
      \dot{\mathbf{\mathfrak{r}}}^i=&\,-(\rho^i+1)\,(\nabla f^i(\mathbf{\mathfrak{r}}^i+\bunderline{\vect{\chi}}^i+\xsi)-\nabla f^i(\xsi))\,+\,\nonumber\\&\sum\nolimits_{k\in\mathcal{T}^i}\big(-\rho^i[\vectsf{w}^i]_{k}^{\top}[\vectsf{w}^i]_{k}\mathbf{\mathfrak{r}}^i
      \,-(\rho^i+1)[\vectsf{w}^{i}]^\top_k\mathfrak{p}_k^i
      \nonumber\\&+\rho^i{[\vectsf{w}}^{i}]^\top_k[\rR_k\,\mathbf{\mathfrak{q}}_k]^i\big).
  \end{align}
  \end{subequations}
Next, consider the Lyapunov function~\eqref{eq::Lya_func} where $(\{\bar{\vect{q}}_k\}_{k=1}^p,\{\vect{p}_k\}_{k=1}^p,\{\vect{\chi}^i\}_{i=1}^N)$ is substituted by $(\{\mathbf{\mathfrak{q}}_k\}_{k=1}^p,\\ \{\mathbf{\mathfrak{p}}_k\}_{k=1}^p,\{\mathbf{\mathfrak{r}}^i\}_{i=1}^N)$.  Following the same argument used to show $\dot{V}\leq0$ in~\eqref{eq::dot_V_CG}, we can show that the derivative of $V(\{\mathbf{\mathfrak{q}}_k\}_{k=1}^p,\{\mathbf{\mathfrak{p}}_k\}_{k=1}^p,\{\mathbf{\mathfrak{r}}^i\}_{i=1}^N)$ along the trajectories of~\eqref{eq::Alg_col_q2}-\eqref{eq::Alg_col_chi}, when~\eqref{eq::Th1_q1} holds, is also negative semi-definite.
 Thus, any equilibrium point $(\{\bunderline{\bar{\vect{q}}}_k\}_{k=1}^p,\{\bunderline{\vect{p}}_k\},\{\bunderline{\vect{\chi}}^i\}_{i=1}^N)$ of~\eqref{eq::Alg_col_q2}-\eqref{eq::Alg_col_chi} is Lyapunov stable (recall~\eqref{eq::Th1_q1}). Therefore, since the trajectories of~\eqref{eq::Alg_col_q2}-\eqref{eq::Alg_col_chi} are approaching to the set of stable equilibrium points,  starting from any initial condition,
the trajectories of~\eqref{eq::Alg_col_q2}-\eqref{eq::Alg_col_chi} converge to a point in its equilibrium set. Consequently, given the change of variables~\eqref{eq::new_coord}, we conclude that starting from stated initial conditions in the statement, the trajectories of~\eqref{eq::Alg} converge, as $t\to\infty$, to a point in its set of equilibrium points~\eqref{eq::eqillibria}, where $(\{\dot{v}^l_k\}_{l\in\VV_k}=\vect{0},\{\dot{y}^l_k\}_{l\in\VV_k}=\vect{0},\{\dvect{x}^i\}_{i=1}^N=\vect{0})$. Therefore, under the stated initial condition, as $t\to\infty$, the limit  point $(\{{v}^l_k\}_{k=1}^p,\{{y}^l_k\}_{k=1}^p,\{\vect{x}^i\}_{i=1}^N)$, $i\in\VV$, $l\in\VV_k$
that satisfies $(\{\dot{v}^l_k\}_{l\in\VV_k}=\vect{0},\{\dot{y}^l_k\}_{l\in\VV_k}=\vect{0},\{\dvect{x}^i\}_{i=1}^N=\vect{0})$ in~\eqref{eq::Alg} is equal to $(\nu^\star_k\vect{1}_{N_k},\vectsf{y}^\star,\{\vectsf{x}^{i\star}\}_{i=1}^N)$, where $(\{\nu^\star_k\}_{k=1}^p,\vectsf{x}^{i\star})$, where $(\{{\nu}_k^\star\}_{k=1}^p,\{{\vectsf{x}^{i\star}}\}_{i=1}^N)$ is a point satisfying the KKT conditions~\eqref{eq::KKT_1}  of problem~\eqref{eq::prob_def-aug2} (this point is not necessarily the point used in the change of variable~\eqref{eq::new_coord}).
 \end{pf}

  \vspace{-0.05in}
\begin{pf}[Proof of Theorem~\ref{prop::main-exp}]
Follow the proof of Theorem~\ref{thm::main} until the choice of the candidate Lyapounv function where we use the candidate function below consisted of $V$ in~\eqref{eq::Lya_func} plus an extra positive quadratic term 
  \begin{align*}
  &\bar{V}(\{\bar{\vect{q}}_k\}_{k=1}^p,\{\vect{p}_k\}_{k=1}^p,\{\vect{\chi}^i\}_{i=1}^N)=V+\\&\quad  \sum\nolimits^{p}_{k=1}\frac{\phi_k}{2}(\vect{\chi}_k+\vectsf{\psi}_k^{\top}\vect{\Gamma}_k\vect{p}_k)^\top\!
   (\vect{\chi}_k\!+\!\vectsf{\psi}_k^{\top}\vect{\Gamma}_k\vect{p}_k)=\vect{\zeta}^\top\vect{E}\vect{\zeta},\end{align*}
  where $\phi_k\in\realpositive$ satisfies 
  $
    \phi_k<\min\{\frac{2(1+\underline{\rho})m}{p(M^2(\bar{\rho}^2+1)^2\!+1)},\\ \frac{2\beta_k\lambda_{2k}}{(\beta^2_k\lambda_{Nk}^2\bar{\rho}^2+\bar{\rho}+1)\|\vect{\psi}_k\|^2}\},
  $
  with  $\underline{\rho}=\min\{\rho^i\}_{i=1}^N$ and $\bar{\rho}=\max\{\rho^i\}_{i=1}^N$.
Here $\vect{\zeta}=[\{\bar{\vect{q}}_k^{\top}\}_{k=1}^p,\{\vect{p}_k^\top\}_{k=1}^p,\{\vect{\chi}^{i\top}\}_{i=1}^N]^\top$ and $\vect{E}>0$ is the
obvious matrix describing the coefficients of the quadratic terms of $\bar{V}$. When every $\GG_k$, $k\in\mathbb{Z}_1^p$ is a connected graph, $\bar{V}$ is a radially unbounded and positive definite function.
Then,
\begin{align*}
\dot{\bar{V}}=&-\!\sum\nolimits_{i=1}^N(\rho^i+1){\vect{\chi}^i}^\top\vect{h}(\vect{\chi}^i)\!+\!\sum\nolimits_{k=1}^p\Big(\!-\beta_k\vect{p}_k^\top\lL_k\vect{p}_k\nonumber\\
      &\,-(\vectsf{\psi}_k\,\vect{\chi}_k-\rR_k\bar{\vect{q}}_k)^\top\vect{\Gamma}_k(\vectsf{\psi}_k\,\vect{\chi}_k-\rR_k\bar{\vect{q}}_k)\\&
      -\frac{\phi_k}{2}\|\vectsf{\psi}_k^{\top}\vect{\Gamma}_k\vect{p}_k\!+\!(\vect{\Gamma}_k\!+\!\vectsf{I})\vect{h}(\vect{\chi}_k)\|^2+\frac{\phi_k}{2}\vect{\chi}_k^\top\vect{\chi}_k\\
      &-\frac{\phi_k}{2}\|\vect{\chi}_k+\beta_k\vectsf{\psi}_k^{\top}\vect{\Gamma}_k\lL_k\vect{p}_k+\vectsf{\psi}_k^{\top}(\vect{\Gamma}_k+\vectsf{I})\vect{p}_k)\|^2\\
&-\phi_k\vect{\chi}^\top_k(\vect{\Gamma_k}+\vectsf{I})\vect{h}(\vect{\chi}_k)+\frac{\phi_k}{2}\vect{h}^\top(\vect{\chi}_k)(\vect{\Gamma}_k+\vectsf{I})^2\vect{h}(\vect{\chi}_k)\\&
+\frac{\beta^2\phi_k}{2}\|\vect{p}_k^\top\lL_k\vect{\Gamma}_k\vectsf{\psi}_k\|^2
+\frac{\phi_k}{2}\vect{p}_k^\top\vectsf{\psi}_k(\vect{\Gamma}_k+\vectsf{I})\vectsf{\psi}_k^{\top}\vect{p}_k\\
&-\frac{\beta\phi_k}{2}\vect{p}_k^\top(\vect{\Gamma}_k+\vectsf{I})\vectsf{\psi}_k\vectsf{\psi}_k^{\top}\vect{\Gamma}_k\lL_k\vect{p}_k\Big),
\end{align*}
where $\vect{h}(\vect{\chi}_k)=\nabla f(\vect{\chi}_k+\xs_k)-\nabla f(\xs_k)$. When $\rho^i\in\real_{>0}$ for all $i\in\VV$, we can write
\begin{align*}
  &\dot{\bar{V}}\leq -\,(1+\underline{\rho})m\,\vect{\chi}^\top\vect{\chi}+\sum\nolimits_{k=1}^p\Big(-\,\beta_k\,\lambda_{2k}\vect{p}_k^\top\vect{p}_k-\\
  &(\vectsf{\psi}_k\,\vect{\chi}_k\!-\!\rR_k\bar{\vect{q}}_k)^\top\vect{\Gamma}_k(\vectsf{\psi}_k\vect{\chi}_k\!-\!\rR_k\bar{\vect{q}}_k)\!+\!
  \!\frac{\phi_k}{2}(M^2(\bar{\rho}+1)^2\!+\!1)\\&\vect{\chi}^\top\!\vect{\chi}\!
  +\frac{\phi_k}{2}(\beta_k^2\lambda_{Nk}^2\bar{\rho}^2\!+\!\bar{\rho}+1)\|\vectsf{\psi}_k\|^2\vect{p}_k^\top\vect{p}_k\!\Big).
\end{align*}
Here, we used the $M^i_l$-Lipschitzness property of local gradients to write  $\vect{h}(\vect{\chi}_k)^\top(\vect{\Gamma}_k+\vectsf{I})^2\vect{h}(\vect{\chi}_k)\leq \sum\nolimits_{i=1}^{N_k} (\rho^i+1)^2\,M^2 \chi^i\,\!^2\leq M^2(\bar{\rho}+1)^2\vect{\chi}^\top\vect{\chi}$ . We also used  $-\sum_{i=1}^N(\rho^i+1)\vect{\chi}_i^\top\vect{h}(\vect{\chi}_i)\leq  -m(\underline{\rho}+1)\vect{\chi}^\top\vect{\chi}$ due to the $m^i_l$-strong convexity of local cost function $f^i_l$, and $-\vect{p}^\top_k\lL_k \vect{p}_k\leq\vect{0}$, which is true because every $\GG_k$, $k\in\mathbb{Z}_1^p$ is a connected graph. We also used $\|\vect{p}_k^\top\lL_k\vect{\Gamma}_k\vectsf{\psi}_k\|^2\leq\lambda_{Nk}^2\bar{\rho}^2\|\vect{\psi}_k\|^2\vect{p}_k^\top\vect{p}_k$ where $\lambda_{Nk}$ is the maximum eigenlavue of $\lL_k$. We note that for $0<\phi_k<\min\{\frac{2(1+\underline{\rho})m}{p(M^2(\bar{\rho}^2+1)^2+1)},\frac{2\beta_k\lambda_{2k}}{(\beta^2_k\lambda_{Nk}^2\bar{\rho}^2+\bar{\rho}+1)\|\vect{\psi}_k\|^2}\}$, we have $\dot{\bar{V}}<0$. Next, note that we can bound $\dot{\bar{V}}$ by a negative definite quadratic upper bound as
 \begin{align}
    &\dot{\bar{V}}\leq -\,\big((1+\underline{\rho})m-\frac{p\phi_k}{2}(M^2(\bar{\rho}+1)^2\!1\big)\,\vect{\chi}^\top\vect{\chi}+\\
    &\sum\nolimits_{k=1}^p\Big(-(\beta_k\,\lambda_{2k}-\frac{\phi_k}{2}(\beta_k^2\lambda_N^2\bar{\rho}^2\!+\!\bar{\rho}\!+\!1)\|\vectsf{\psi}_k\|^2)\vect{p}_k^\top\vect{p}_k\nonumber\\
  &-(\vectsf{\psi}_k\,\vect{\chi}_k-\rR_k\bar{\vect{q}}_k)^\top\vect{\Gamma}_k(\vectsf{\psi}_k\vect{\chi}_k-\rR_k\bar{\vect{q}}_k)\Big)= -\vect{\zeta}^\top\vect{F}\vect{\zeta},\nonumber \end{align}
 where $\vect{F}>0$ is the obvious matrix describing the coefficients of the quadratic terms of the upper bound of  $\dot{\bar{V}}$. 
 Because $\bar{V}$ is a quadratic positive definite function and the upper bound on $\dot{\bar{V}}$ is a quadratic negative definite quadratic function, by virtue of~\cite[Theorem 4.10]{HKK:02},~\eqref{eq::Alg_col_q2}-\eqref{eq::Alg_col_chi} is exponentially stable, and its trajectories converge to the origin with the rate no worse than $\frac{\lambda_{\min}(\vect{F})}{2\lambda_{\max}(\vect{E})}$, where $\lambda_{\min}(\vect{F})$ is the minimum eigenvalue of $\vect{F}$ and $\lambda_{\max}(\vect{E})$  is the maximum eigenvalue of $\vect{E}$. Consequently,  starting from any initial condition given in the statement, the trajectories $t\mapsto(\{\vect{v}_k(t)\}_{k=1}^p,\{\vect{x}^i(t)\}_{i=1}^N)$ converge exponentially fast with the rate given above to $(\,{{\nu}}_k^\star\vect{1}_{N_k},\{{\vectsf{x}^{i\star}}\}_{i=1}^N)$, as~$t\to\infty$. 

  \vspace{-0.15in}
If $\rho^i\!=\!0$ for any $i\in\VV$, we can only guarantee that 
$\dot{\bar{V}}\leq0$ with 
$\mathcal{S}=\{(\{\bar{\vect{q}}_k\}_{k=1}^p,\{\vect{p}_k\}_{k=1}^p,\{\vect{\chi}^i\}_{i=1}^N)\in\,\prod_{k=1}^p\real^{N_k-1}\times\,\prod_{k=1}^p{\real^{N_k}~\times}\,\,\, \prod_{i=1}^N\reals^{n^i} |\,
  \dot{\bar{V}}\equiv0\}=
  \Big\{\!\{\bar{\vect{q}}_k\}_{k=1}^p,\!\{\vect{p}_k\}_{k=1}^p,\!\{\vect{\chi}^i\}_{i=1}^N\!\in\!\prod_{k=1}^p\real^{N_k-1}\!\times\!\prod_{k=1}^p\real^{N_k}\times\prod_{i=1}^N\reals^{n^i}\Big|
 ~\vect{p}_k=\vect{0},~\vect{\chi}^i=\vect{0}, \vect{\Gamma}_k\rR_k\bar{\vect{q}}_k=\vect{0},~ i\in\VV, k\in\mathbb{Z}_1^p\}$. Next, we note that since $\rR_k$ is a full column rank matrix, given~\eqref{eq::Th1_q1}, the only trajectory $t
  \mapsto (\{\bar{\vect{q}}_k(t)\}_{k=1}^p,\{\vect{p}_k(t)\}_{k=1}^p,\{\vect{\chi}^i(t)\}_{i=1}^N)$ of~\eqref{eq::Alg_col_q2}-\eqref{eq::Alg_col_chi} that belongs to
  $\mathcal{S}$ for all $t\in\real_{\geq 0}$ is  $(\{\bar{\vect{q}}_k(t)\equiv \vect{0}\}_{k=1}^p,\{\vect{p}_k(t)\equiv\vect{0}\}_{k=1}^p,\{\vect{\chi}^i(t)\equiv\vect{0}\}_{i=1}^N)$. Therefore, using a LaSalle invariant set analysis of~\cite[Corollary 4.1]{HKK:02}, and recalling the change of variable~\eqref{eq::new_coord} and also \eqref{eq::Th1_q1}, we can conclude that $t\mapsto(\{\vect{v}_k(t)\}_{k=1}^p,\{\vect{x}^i(t)\}_{i=1}^N)$ of~\eqref{eq::Alg} converges exponentially fast to $(\,{{\nu}}_k^\star\vect{1}_{N_k},\{{\vectsf{x}^{i\star}}\}_{i=1}^N)$.
\end{pf}

\vspace{-0.25in}

\section*{Appendix B}
\renewcommand{\theequation}{B.\arabic{equation}}
\setcounter{equation}{0}
\noindent Consider the optimization problem
\begin{align}\label{eq::example3}
\!\! \xs\!=\arg\min_{\vect{x}\in
    \reals^{2}} \sum\nolimits_{i=1}^2 f^i(x^i)\,~\text{subject~to~}\,x^1+x^2=2,
\end{align}
where $f^i(x^i)\!=\!\begin{cases}0,& |x^i|\leq 2,\\
\,\frac{1}{2\alpha}(|x^{i}|-2)^2,& 2<|x^i|\leq 2+\alpha,\\
(|x^i|-2-\frac{1}{2}\alpha),& |x^i|> 2+\alpha,
\end{cases}$\\
with $\alpha\!=\!0.01$. Here, the cost function is convex.

\begin{figure}[!htb]
  \unitlength=0.5in
  \centering 
    \includegraphics[scale=0.5]{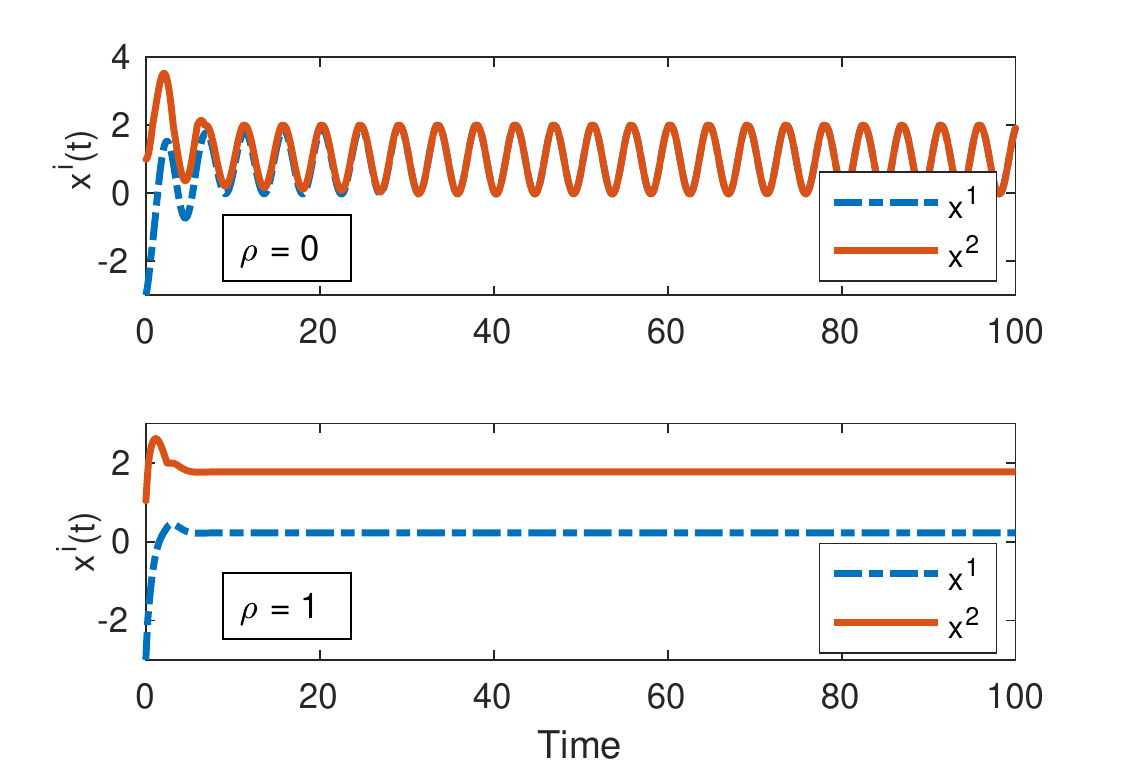}
   \caption{\scriptsize Trajectories of algorithm~\eqref{eq::saddle-aug} when it is used to solve  optimization problem~\eqref{eq::example3} with $\rho=0$ and $\rho=1$.
   }\label{fig::Ex3}
\end{figure}

Note that the optimization problem~\eqref{eq::example3} has infinite number of minimizers that correspond to the minimum cost of $f^\star=0$. One of these minimizers is $(x^{1\star},x^{2\star})=(0,2)$. Figure~\ref{fig::Ex3} shows the $x^i$ trajectories of central solver~\eqref{eq::saddle-aug} over time. As shown, the algorithm does not converge when $\rho=0$, while the convergence is achieved when we use the augmented Lagrangian with $\rho=1$.

\end{document}